\documentclass[twocolumn, aps, superscriptaddress]{revtex4-1}

\usepackage{graphicx}
\usepackage[caption=false]{subfig}
\usepackage{amsmath} 
\usepackage[shortlabels]{enumitem}
\usepackage{upgreek}

\renewcommand{\vec}[1]{\boldsymbol{#1}}
\newcommand{\tens}[1]{\boldsymbol{#1}}
\newcommand{\bnabla}{\vec{\nabla}}
\newcommand{\unitvec}[1]{\boldsymbol{\hat{#1}}}
\newcommand{\fig}[1]{Fig.~\ref{#1}}
\newcommand{\mov}[2]{Movie~S{#1}}
\newcommand{\eq}[1]{Eq.~\ref{#1}}
\newcommand{\mthds}[1]{Materials and Methods}
\newcommand{\sifig}[2]{Fig.~S{#1}}
\newcommand{\simov}[1]{Movie~S{#1}}

\begin{document}


\title{Submersed Micropatterned Structures Control Active Nematic\\Flow, Topology and Concentration}


\author{Kristian Thijssen}
  \thanks{These authors contributed equally}
  \affiliation{The Rudolf Peierls Centre for Theoretical Physics, Department of Physics, University of Oxford, 1 Keble Road, Oxford OX1 3PU, UK}
\author{Dimitrius Khaladj}
  \thanks{These authors contributed equally}
  \affiliation{Department of Physics, University of California, Merced, CA, 95343, USA}
\author{S. Ali Aghvami}
    \affiliation{Physics Department, Brandeis University, Waltham, MA, 02453, USA}
  \affiliation{Department of Molecular Metabolism, Harvard University, Boston, MA, 02115, USA}
\author{Mohamed Amine Gharbi}
  \affiliation{Department of Physics, University of Massachusetts Boston, Boston, MA, 02125, USA}
\author{Seth Fraden}
  \affiliation{Physics Department, Brandeis University, Waltham, MA, 02453, USA}
\author{Julia M. Yeomans}
  \affiliation{The Rudolf Peierls Centre for Theoretical Physics, Department of Physics, University of Oxford, 1 Keble Road, Oxford OX1 3NP, UK}
\author{Linda S. Hirst}
  \email{lhirst@ucmerced.edu}
  \affiliation{Department of Physics, University of California, Merced, CA, 95343, USA}
\author{Tyler N. Shendruk}
  \email{t.shendruk@ed.ac.uk}
  \affiliation{SUPA, School of Physics and Astronomy, The University of Edinburgh, Peter Guthrie Tait Road, Edinburgh EH9 3FD, United Kingdom}


\begin{abstract}
\noindent 
Coupling between flows and material properties imbues rheological matter with its wide-ranging applicability, hence the excitement for harnessing the rheology of active fluids for which internal structure and continuous energy injection lead to spontaneous flows and complex, out-of-equilibrium dynamics. We propose and demonstrate a convenient, highly tuneable method for controlling flow, topology and composition within active films. Our approach establishes rheological coupling via the indirect presence of fully submersed micropatterned structures within a thin, underlying oil layer. Simulations reveal that micropatterned structures produce effective virtual boundaries within the superjacent active nematic film due to differences in viscous dissipation as a function of depth. This accessible method of applying position-dependent, effective dissipation to the active films presents a non-intrusive pathway for engineering active microfluidic systems.
\end{abstract}

\maketitle

Active fluids are inherently out-of-equilibrium; they locally transform internal energy into material stresses that can result in spontaneous, hydrodynamic motion. 
An increasing number of biophysical systems, including colonies of bacilliform microbes\cite{dell2018,You2018,Li2019,echten2020}, cellular monolayers\cite{Duclos2017,Saw2017,Kawaguchi2017,Duclos2018,Perez2019}, and subcellular filaments\cite{Zhang2017,Huber2018,Kumar2018} display such collective active motion, orientational order and topological singularities. 
Controlling active dynamics is essential not only to fully understanding how such biological systems employ self-generated stresses but also in order to develop active-microfluidic devices. 

To this end, recent work considers how confining walls\cite{Norton2018,Opathalage2019,Hardouin2020}, arrangements of obstacles\cite{Reinken2020,Zhang2020} and the dynamics of topological defects\cite{Tan2019} dictate active nematic flow. 
Control of active material concentration has been studied from the perspectives of co-existence of phases in self-propelled rods\cite{Bertin2015,Nagel2020,Bar2020} and motility-induced phase separation\cite{Linden2019,Caprini2020,Grossmann2020}. 
Controlled accumulation and depletion of active matter has been engineered in bacterial systems to concentrate cells\cite{Mahmud2009,Katuri2018} and to drive bacterial-ratchet motors\cite{Leonardo2009,Sokolov2010,Pietzonka2019}. 
Similarly, substrate gradients modify cellular motility driving density variation\cite{Sunyer2020} and directed migration\cite{Novikova2017,Schakenraad2020}. 

In addition to varying concentration and flow, topology has been controlled by including externally driven flows\cite{Sokolov2019,Norton2020,Rivas2020} and curvature\cite{Ellis2018,Pearce2019}. 
Recent work shows that locally altering activity modifies defect populations\cite{Zhang2021,Shankar2019,tang2021,mozaffari2021}, and anisotropic smectic sublayers below active nematic sheets can constrain orientation\cite{Guillamat2017}. 
Such studies demonstrate how underlying sublayer properties have pronounced effects on active dynamics and suggest approaches for engineering control of active matter. 

\begin{figure*}[tb]
    \centering
    \includegraphics[width=0.8\textwidth]{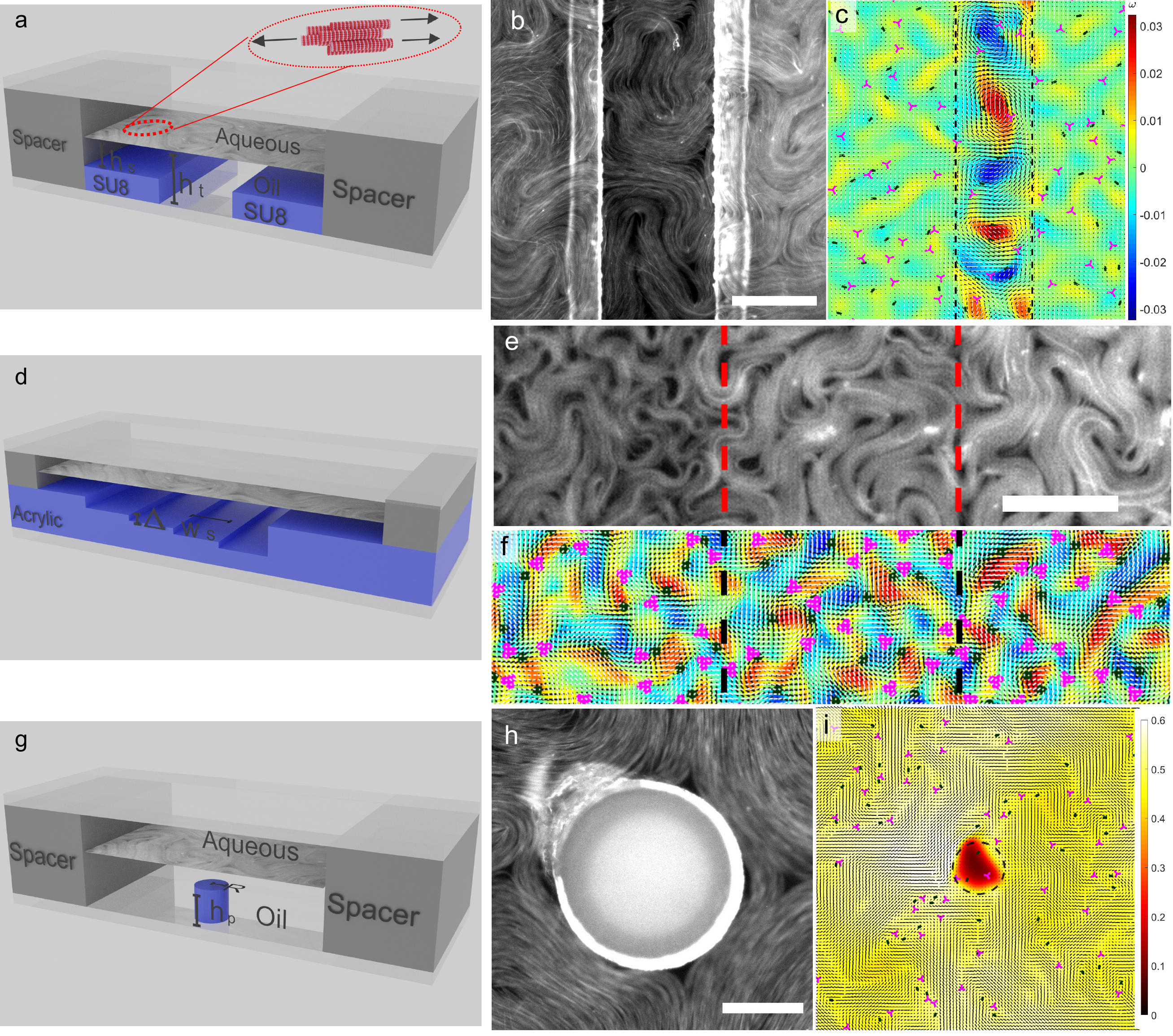}
    \caption{\small
	\textbf{Submersed micropatterns control active nematic dynamics.} 
	(a-c) \textbf{Trench set-up.} 
	An active film resides at the oil-water interface above different substrate depths. 
	The active flows drag the underlying oil layer but viscous dissipation is depth dependent, affecting active nematic film dynamics. 
	(b) Fluorescence microscope image of the active nematic bundled-microtubule film above a submersed trench. 
	Scale bar $= 250 \upmu\textrm{m}$.
	(c) Simulation results for vorticity field within the superjacent active nematic layer. 
	The flow behaviors within the low-friction region (between the dashed lines) are distinct from the behaviour in the high friction region (beyond the dashed lines).  
	Plus-half (minus-half) defects denoted by dark green (magenta) symbols behave differently in the two regions. 
	(d-f) \textbf{Stairway set-up.} 
	(e) Fluorescence microscope image of micro-milled stairway and the superjacent bundled-microtubule film. Scale bar $= 250 \upmu\textrm{m}$. 
	Step location indicated by dashed lines. The oil depth increases from left to right. 
	The differences in oil depth alter the length scale of the active turbulence above each step. 
	(f) Simulations results for discrete steps in the effective friction (dashed lines). 
	The effective friction coefficient decreases from left to right. Colourbar is shared with (c).
	(g-i) \textbf{Pillar set-up.} 
	(h) Fluorescence microscope image of the bundled-microtubule film above the SU-8 micropillar. Scale bar $= 100 \upmu\textrm{m}$. 
	(i) Simulation results show the active nematic concentration $\phi$ is depleted within the high friction region encircled by the pillar perimeter (dashed line). 
    }
    \label{fig:Set_Up}
\end{figure*}

We propose a new micropattern-based method for controlling active nematic dynamics without contiguous contact with active films. 
By patterning oil-submersed solid substrates below 2D active nematic films with geometrical structures of differing height, we achieve effective virtual boundaries within active films that control topological defect populations, collective flow and concentration of active nematic material without penetrating the film. 
By implementing underlying submersed patterned microstructures, we tune the depth of the oil layer to adjust dissipation within the superjacent film and thereby generate a highly tuneable technique for controlling the active dynamics.
Presently, we introduce four initial submersed structures: micropatterned trenches (\fig{fig:Set_Up}a-c), undulated substrates (\sifig{1}{sifig:sinusoid}), stairways (\fig{fig:Set_Up}d-f), and pillars (\fig{fig:Set_Up}g-i).

\subsection*{Trench}

To investigate how structures fully submersed in a layer of oil influence defect dynamics in the superjacent active film, we consider the trench geometry depicted in \fig{fig:Set_Up}a. 
An active nematic microtubule network is generated at an oil-water interface above a micropatterned trench of depth $\Delta_t = 18 \pm 1 \upmu\textrm{m}$ and width $w_t=327 \pm 2 \upmu\textrm{m}$ fabricated using photolithography (\mthds{methods:photo}).
We observe that flows in the active nematic layer exhibit coexistence of two distinct regions: one directly above the trench and another in the shallows surrounding the trench (\fig{fig:Set_Up}b). 
These regions are separated by well-defined virtual boundary lines located directly above the trench edges. 
The trench edges are visible as a pair of parallel white lines due to stress-induced auto-fluorescence of the micropatterned photoresist in regions of precipitous height change.
Beyond the trench boundaries, the active nematic retains the chaotic nature of active turbulence; however, within the boundaries, the trench width establishes a local confining length scale within the superjacent active nematic film (\mov{1}{mov:trenchExp}).
These virtual walls trap defects in the trench region and produce active flow behaviours comparable to those observed in confining channels\cite{Shendruk2017,Doostmohammadi2017,Chandragiri2019,Opathalage2019,Hardouin2019}. 
This result illustrates how such effective virtual boundaries can be used to define areas of orderly flows and areas of active turbulence without penetrating the active film.

\begin{figure}
    \centering
    \includegraphics[width=0.475\textwidth]{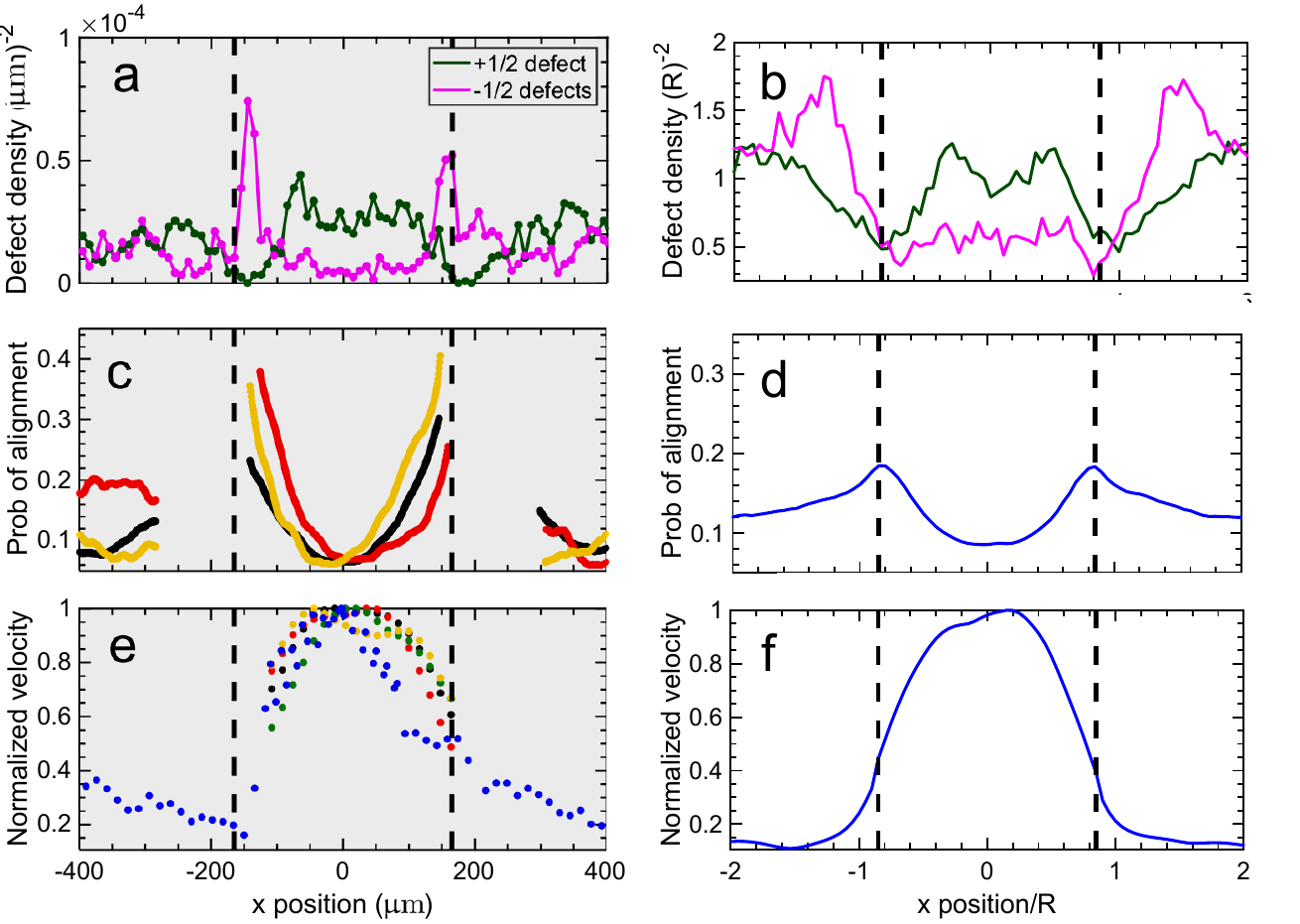}
    \caption{\small
	\textbf{Positive defects depleted at the trench interface }
	Panels with shaded grey backgrounds denote experimental results, while non-shaded backgrounds denote numerical results. 
	(a-b) Distribution of $+1/2$ (dark green) and $-1/2$ (magenta) defects as a function of transverse position $x$ measured from experiments and simulations respectively. 
	(c-d) The probability profile that the nematic director is oriented less than $10^\circ$ from the direction parallel to the trench wall as a function of $x$. 
	The director field has a high probability of alignment with the virtual boundary. 
	The gap in data at the trench edge is explained in \mthds{methods:photo}. 
	(e-f) Normalised root mean square fluid velocity profile across the trench. Experimental results in (e) are calculated from two different identical trenches. The experimentally measured maxima are $\{4.00,1.68,1.57,1.00,1.12\}\upmu\textrm{m}/s$. Colours in (c) and (e) denote different experimental iterations.
    }
    \label{fig:Defect_depletion}
\end{figure}

The $\pm$1/2 topological defect distributions across the trench (\mthds{methods:data}) demonstrate that $-1/2$ defects tend to be located in the vicinity of the virtual boundary (\fig{fig:Defect_depletion}a).
Experimental observations of $-1/2$ defect trajectories near the boundary reveal that they tend to linger over long intervals, contributing to peaks (\mov{1}{mov:trenchExp}). 
In contrast, $+1/2$ defects tend to be depleted from the vicinity of the trench boundary and are confined within the trench region, moving along oscillatory trajectories that do not typically approach the boundaries (\mov{1}{mov:trenchExp}). 
In the exterior region, far from the virtual trench, the defect density profile approaches a homogeneous distribution of positive and negative defects. 

The effective virtual boundaries arise from abrupt steps in fluid depth $h\left(\vec{r}\right)$ between the film and the underlying substrate at each point $\vec{r}$. 
The fluid depth $h$ increases from $h_s$ in the surrounding shallows to a trench depth $h_t=h_s+\Delta_t$ (\fig{fig:Set_Up}a). 
As activity drives flows within the nematic film, the underlying oil layer viscously dissipates momentum due to the subjacent no-slip substrate, which can be described as a local effective friction $\gamma(\vec{r})$ acting on each point within the superjacent active film\cite{pismen2017}. 
Following from the lubrication limit, the effective friction coefficient scales as $\gamma \sim \eta'/h\left(\vec{r}\right)$, where $\eta'$ is an effective viscosity of the film and surrounding fluids. 
The abrupt height change across the trench boundaries results in sharp, virtual boundaries. 

We replicate the observed experimental phenomena with 2D active nemato-hydrodynamic simulations, in which the submersed micropatterns are incorporated via an effective friction field (\mthds{methods:sims}). 
Numerical results demonstrate that a step in effective friction can reproduce the experimentally observed active flows (\fig{fig:Set_Up}c) and introduce virtual boundaries in the active layer, which repel $+1/2$ defects (\fig{fig:Defect_depletion}a-b). 
Integrating the defect density across the channel in \fig{fig:Defect_depletion}b gives net zero charge; however, it is more challenging to accurately identify minus defects than positive in experiments. As a result, this labeling bias causes a slight net positive charge in \fig{fig:Defect_depletion}a.
The qualitative agreement in behaviour between experimental and numerical defect distributions demonstrates how effective friction is the mechanism by which micropatterned structures create virtual planar boundaries and introduce a confinement length scale to the active nematic without penetrating the film. 
The $-1/2$ defect density peak at the virtual boundary (\fig{fig:Defect_depletion}a) is consistent with work\cite{Opathalage2019,Hardouin2019} showing walls can act as catalysts for pair creation and unbinding: while newly created $+1/2$ defects move away due to self-propulsion, the $-1/2$ defects remain near the boundary. 

While the experimental $-1/2$ defect density peaks sharply in the vicinity of the virtual boundaries (\fig{fig:Defect_depletion}a), it is broadened and peaked outside the trench region in simulations (\fig{fig:Defect_depletion}b). 
To understand this difference, we consider the time-averaged director orientation across the trench (\mthds{methods:data}; \fig{fig:Defect_depletion}c), which reveals that the virtual boundaries introduce an effective alignment of the director as the probability of aligning to the interface becomes different from the bulk, similar to that seen for impermeable boundaries\cite{Opathalage2019,Hardouin2019}. 
This is because any orthogonal bundle midway over the boundary is subject to a large axial laminar flow inside and slow disorderly flows outside, which compete to produce an aligning torque. 
The model captures this behaviour, showing that the probability declines to a uniform distribution far from the trench (\fig{fig:Defect_depletion}d). 
Experiments exhibit stronger planar alignment at the virtual boundaries than simulations. This is likely related to the model’s assumption of a continuous fluid (\mthds{methods:sims}), which is in contrast to the network of finite-sized microtubule bundles that act as material lines preventing defects from crossing\cite{Opathalage2019}, resulting in an accumulation.

This is reminiscent of simulation studies of passive nematic tactoids where defects travel outside the interface and become virtual unless an anchoring term is included\cite{metselaar2017electric}. 
Future research is needed to more fully understand if an effective term can be devised to account for this effect in continuum models that do not contain sharp $\phi$ interfaces. 
The stronger alignment in the experiments constrains the $-1/2$ defects to the region inside the trench, while in simulations they are pushed to the outside of the boundaries (\fig{fig:Defect_depletion}a-b). 
In both experiments and simulations, $+1/2$ defects are trapped between the virtual boundaries. 

The submersed trench not only impacts the nematic field but also generates a virtual boundary for the velocity field (\fig{fig:Defect_depletion}e-f). 
Superjacent to the trench, velocities are lower in the proximity of the trench boundary and maximum at the trench centre. 
Since activity varies slightly between experimental realisations, we normalise the flow profiles (\fig{fig:Defect_depletion}e) and compare to the decrease predicted by the model (\fig{fig:Defect_depletion}f). 
The virtual boundaries do not impose no-slip conditions but decrease the speed to the slower value of the surrounding active turbulence, which can be seen explicitly for an experimental instance (blue dots; (\fig{fig:Defect_depletion}e)) and in simulations (\fig{fig:Defect_depletion}f). 
The decreasing flow profile explains the preferential alignment of the microtubule bundles in the vicinity of the virtual boundaries (\fig{fig:Defect_depletion}c-d). 
Any orthogonal bundle midway over the boundary is subject to a large, axial, laminar flow inside and slow disorderly flows outside, which compete to produce an aligning torque. 

Comparing the faster flow profile above the trench to the slower, disorderly active turbulence in the exterior region calls attention to the fact that active turbulence is a low-Reynolds number phenomenon\cite{Alert2020}. 
Confinement screens chaotic flows that would otherwise develop on scales larger than the trench width, while the low friction allows the rapid-but-steady flow profiles superjacent to the trench (\fig{fig:Set_Up}b-c). 
On the other hand, the higher friction produces a smaller characteristic length scale\cite{Thijssen2020}, but also slower speeds in the shallows. 
Thus, the submersed micropatterned trench segregates rapid laminar flow above the trench and slow-but-disorderly active turbulence outside. 

\begin{figure*}
    \centering
    \includegraphics[width=0.8\textwidth]{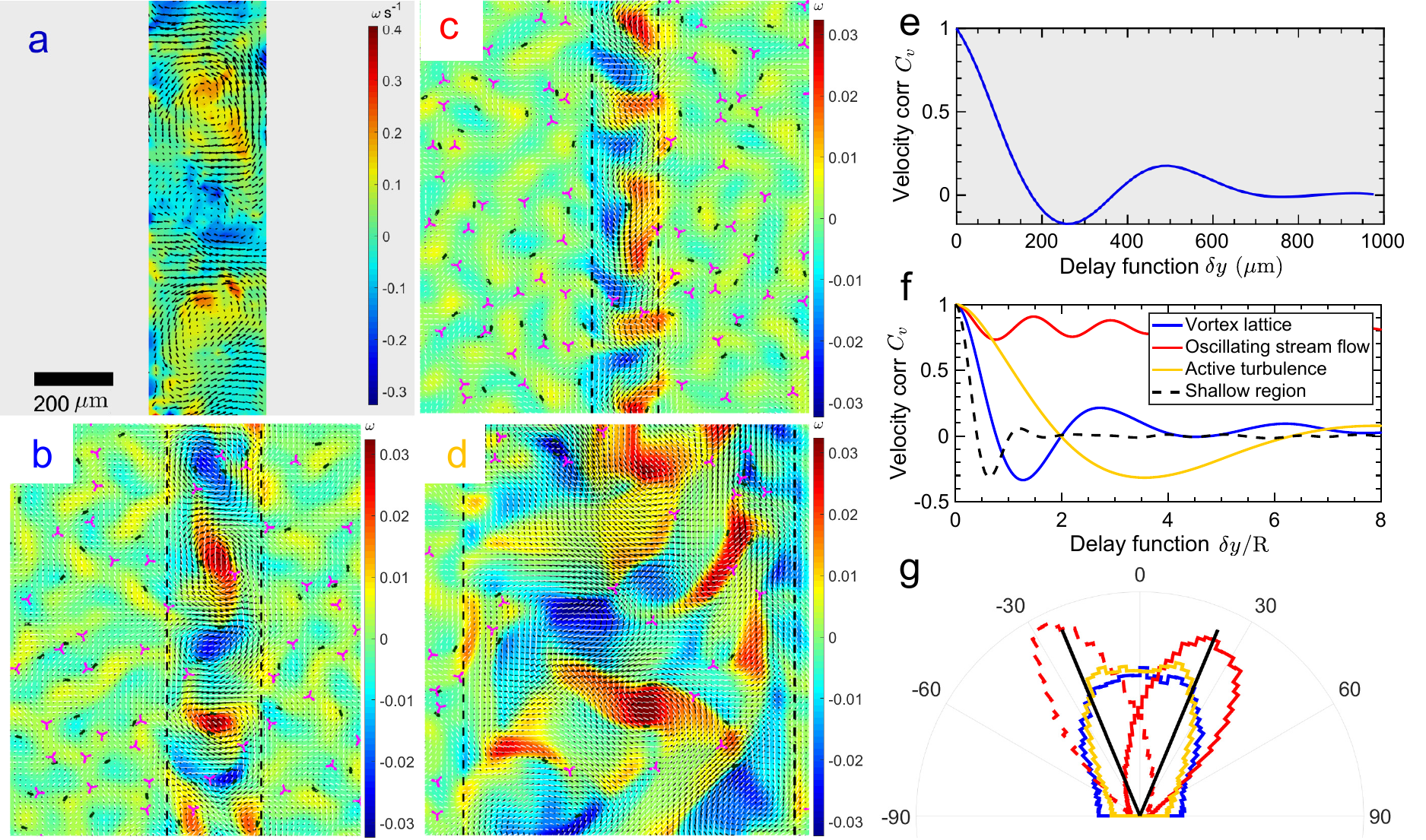}
    \caption{\small
	\textbf{Friction boundaries result in separate flow regions.}
	(a) Instantaneous experimental vorticity superjacent the deep trench region from Particle image velocimetry (PIV) for a trench of width $w_t=325\upmu\textrm{m}$. 
	Panels with shaded grey backgrounds denote experimental results, while non-shaded backgrounds denote numerical results. 
	(b) Simulation snapshot of a repeating lattice of counter-rotating vortices above a trench of width $w_t=1.7R$ (\mthds{methods:sims}). 
	Plus-half defects (dark green) are trapped between the virtual boundaries, generating the repeating vortex structure along the centre line that is distinct from the active turbulence that exists outside the virtual boundaries. 
	(c) Decreasing the trench width to $w_t=1.2R$ in simulations results in long-range, oscillatory, streaming flow inside the trench. 
	(d) Increasing the trench width to $w_t=6R$ in simulations results in active turbulence both inside and outside the trench region, but with differing intrinsic length scales due to the different effective frictions.
	(e) The velocity-velocity autocorrelation function $C_v(\delta y) = \left\langle \vec{v}(\vec{r};t) \cdot \vec{v}(\vec{r}+\delta y \ \unitvec{y};t) \right\rangle / \left\langle v^2 \right\rangle$ 
	for the experiment illustrated in (a), measured a distance $\delta y$ along the $x=0$ centre line of the trench. 
	Due to the confinement effect, long-range flow structures are formed in the low-friction regime epitomised by the strong correlation-anti-correlation-correlation signal. 
	(f) Autocorrelation functions calculated from the simulations inside the channel shown in (b) (blue), (c) (red) and (d) (yellow), and the shallow region outside the virtual channel (dashed).
	The blue curve displays pronounced correlation and anti-correlation indicating the counter-rotating vortex pattern in (b), which corresponds to the behaviour observed in the presented experiments of (a,e). 
	The red curve is long-lived and fully correlated in the narrow channels of (c), while the yellow curve decorrelates to zero after an anti-correlation, signalling active turbulent behaviour both within and outside the virtual boundaries. 	
	(g) The director angle probability density at the friction boundaries of (b-d). The solid (dashed) lines correspond to measurements at the right (left) friction boundary. The black lines are the analytical Leslie angles ($\theta_\text{L}=22^\circ$). 
    }
    \label{fig:dif_flows}
\end{figure*} 

Because the submersed micropatterned trench produces virtual boundaries that introduce a confining length scale, the competition between confinement and intrinsic active nematic length scale can be probed. 
\mov{1}{mov:trenchExp} and \fig{fig:dif_flows}a experimentally demonstrate that a recurrent vorticity structure is established between the virtual boundaries when active and confining scales coincide\cite{Shendruk2017,Opathalage2019,Zhang2020} and simulations underscore the periodicity of counter-rotating vorticies (\fig{fig:dif_flows}b and \mov{2}{mov:trenchSim}). 
Examining the velocity autocorrelation functions quantifies the different flow profiles above the trench (\fig{fig:dif_flows}e and f; blue curve). 
The correlation function exhibits repetition between correlated and anti-correlated regions due to repeating clock-wise and anti-clockwise vortices. 
Active turbulence exists outside of the virtual channel, as characterised by an immediate initial drop in the correlation (\fig{fig:dif_flows}f; dashed curve). 

In narrower or wider confinements, the flow transitions to other states (\fig{fig:dif_flows}c-f). 
In simulations of the narrow trench (\fig{fig:dif_flows}c), the flows are long-ranged and oscillatory with a preference for aligning with the boundaries (\fig{fig:dif_flows}f). 
This oscillatory-streaming state occurs when the confining length scale $w_t$ is small compared to the low-friction intrinsic active nematic length scale\cite{Shendruk2017}. 
Increasing the trench width predicts active turbulence in both the area superjacent to the trench and the shallow exterior regions (\fig{fig:dif_flows}d) but with differing active nematic length scales (\fig{fig:dif_flows}f). 
Equivalently higher levels of activity decrease the active length scale and we have experimentally observed instances of disorderly flow states in both regions. 
The structure of the different flow states impacts the effective director alignment observed in \fig{fig:Defect_depletion}d. 
For streaming flow which is dominated by a simple shearing flow with small fluctuations, we find the most likely angle the director makes with the virtual boundary, as seen in \fig{fig:dif_flows}g. 
These angles are distributed about the Leslie angle $\theta_\text{L}=22^\circ$ (\mthds{methods:sims}), furthering our argument that the preferred orientation is due to flow alignment. 
In the wider trenches, the director varies widely due to defect creation events\cite{Hardouin2019}, but we still see a symmetry breaking as perpendicular alignment to the interface is disfavoured due to the aligning torque. 
The relative position of the $-1/2$ defect peak to the interface does not change as we alter the trench width. 

The trench geometry demonstrates that submersed microstructure patterning can impose confining virtual boundaries and is a feasible technique for maintaining coexistence of distinct flow behaviours simultaneously at different locations in a single active nematic layer. 
A strength of our submersed-micropatterned-structure approach is that the boundaries act without physically penetrating the film and so active material does not first have to saturate a cavity before confinement dynamics can be explored\cite{Opathalage2019}. 
Filling complex geometries with filament-based active material may be the prohibitive step in active microfluidics\cite{Kempf2019}. 
The proposed micropatterned method circumvents these difficulties, opening possibilities for experiments involving more complex geometries and fine-tuned positional control.

\subsection*{Sinusoid Substrates}

While the trench geometry demonstrates that micropatterns with precipitous edges can actualise abrupt boundaries within superjacent active films, the ability to gradually tune the effective friction through gradients in oil-layer thickness allows our technique to gently guide defect dynamics. 
To demonstrate this, an undulating effective friction is produced by fully submersing a micropatterned 1D sinusoidal substrate (\mthds{methods:milling}) characterised by amplitude $\Delta_u = 40 \pm 2 \upmu\textrm{m}$ and wavelength $\lambda_u=150 \pm 2 \upmu\textrm{m}$. 
Unlike the trench geometry, the sinusoid system does not separate into distinct coexisting flow states (\mov{3}{mov:undulSmall}). 
Rather, the resulting anisotropic friction gradients present a means of orientation-control of motile defects. 
Self-propelled $+1/2$ defects orient and travel in trains above the troughs (\sifig{1}{sifig:sinusoid}a). 
Motile defects move through the system subject to friction gradients when they have components perpendicular to the troughs, such that trajectories co-aligned with the troughs minimise dissipation, causing the observed parallel/anti-parallel laning of $+1/2$ defects. 
Similar trains have been observed in active nematic layers above smectics\cite{Guillamat2016} but, since the spacing between smectic layers is significantly smaller, the defect trains in that configuration are not caused by gradients in effective friction; rather, they are due to uniform anisotropic friction. 

However, such trains of +1/2 defects do not persist indefinitely since the trains produce nematically ordered regions, which are susceptible to the extensile-active nematic hydrodynamic-bend instability\cite{Aditi2002,giomi2011,martinez2019} causing pair creation events that inevitably destabilise the flow (\mov{3}{mov:undulSmall}). 
Since initially unbound $+1/2$ defects are typically oriented perpendicular to the nematic ordered lanes (\sifig{1}{sifig:sinusoid}b), this produces a crosshatched trajectory pattern. 
These dynamics are not nearly as pronounced in sinusoid systems with a larger wavelength ($\lambda_u=500 \upmu\textrm{m}$; \mov{4}{mov:undulBig}). 
Positive defects that are partially oriented along the friction gradient exhibit the same deflected motion as in the smaller wavelength system and so the motile defects show some alignment along the troughs but the crosshatched dynamics are indiscernible. 
The sinusoid geometry demonstrates that submersed micropatterned structures can fine-tune the active flow and nematic structure, thereby offering a means to guide and control defect dynamics.

\subsection*{Stairs}

\begin{figure}
    \centering
    \includegraphics[width=0.475\textwidth]{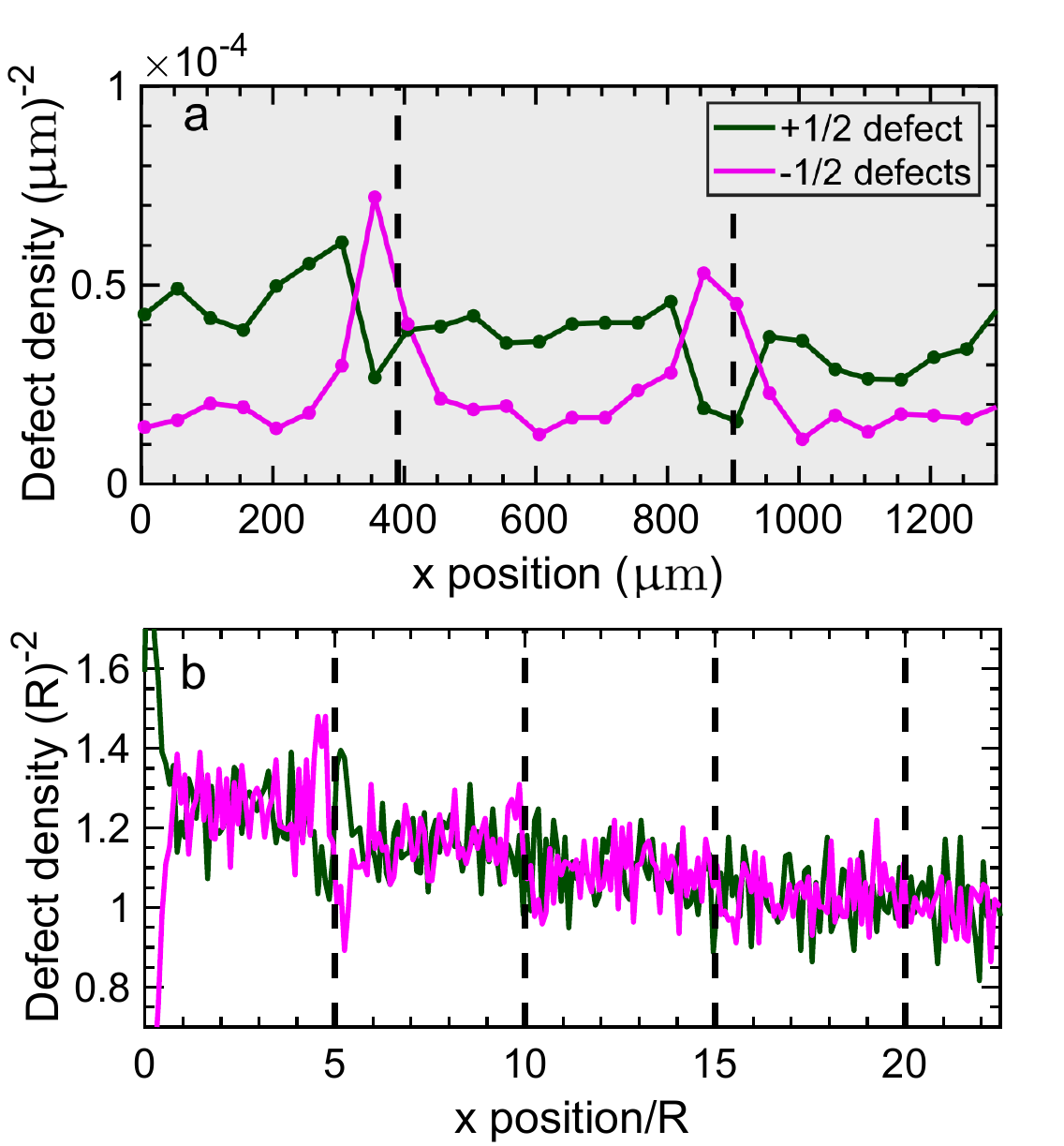}
    \caption{\small
	\textbf{Submersed micropatterned stairway allows simultaneous coexistence of regions of separated active turbulence with differing defect densities, while adenosine triphosphate (ATP) concentration remains constant.} 
	Plus-half (dark green) and minus-half (magenta) defect distribution as a function of position down the stairway $x$. 
	(a) Measurements from experiments for steps of width $w_t=500\upmu\textrm{m}$ (grey shaded background). 
	(b) Measurements from simulations for steps of width  of width $w_t=5R$. 
    }
    \label{fig:stairs}
\end{figure} 

We now present a substrate patterned as a submersed stairway (\fig{fig:Set_Up}d) designed to simultaneously observe active turbulence and gradations in the characteristic length scales (\mov{5}{mov:stairs}). 
Individual steps are micromilled to possess horizontal width $w_s = 500 \pm 2 \upmu\textrm{m}$ and height of $\Delta_s = 10 \pm 1 \upmu\textrm{m}$ (\mthds{methods:milling}). 
The fluid depth is $h\left(\vec{r}\right) = h_0 + \Delta_s n\left(x\right)$, for step number $n$ and initial fluid depth $h_0= 12 \pm 3 \upmu\textrm{m}$, determined through confocal microsopy (\sifig{2}{sifig:SIconfocalmeasurement}; \mthds{methods:milling}). 
We focus on steps $7\leq n \leq 9$ for which the microtubule network forms a well-defined, continuous nematic field (\fig{fig:Set_Up}e; \mov{5}{mov:stairs}). 
As the depth increases with $n$, the effective dissipation within the oil-layer decreases, which we simulate via discrete steps in effective friction in the superjacent active film (\fig{fig:Set_Up}f; \mthds{methods:sims}). 

Above the step pattern, the active length scales increase with decreasing friction, as characterised by the defect distribution (\fig{fig:Set_Up}e-f; \fig{fig:stairs}a; \mov{5}{mov:stairs}). 
Within each step the distribution is flat. However, at each edge, the number density of $-1/2$ defects peaks, while the density of $+1/2$ defects plunges. This is consistent with defect densities at the edges of the trench (\fig{fig:Defect_depletion}a-b). 
Though the simulated edge peaks are less pronounced than in experiments, numerical results show more clearly the decrease in defect density across multiple steps.   
The change in defect density is modest, consistent with studies demonstrating that increasing oil viscosity five orders of magnitude only increases defect density by a factor of order unity\cite{Guillamat2016b,martinez2021}, which highlights the potential tunability of our method. 
Because the effective friction is inversely proportional to oil depth, changes to the defect density become less pronounced with step number and, in the large-oil-depth limit, substrate features become indiscernible within the nematic flows. 

Interestingly, we only observe a continuously well-defined nematic field for oil depths that are much greater than $h_0$ in both experiments and simulations (\fig{fig:Set_Up}e). 
We could reliably resolve the nematic order and measure defect density with low experimental uncertainty for steps $n \geq 7$. 
For small $n$, the active film exhibits disorderly nonhomogeneous textures akin to those observed in experiments utilising high viscosity oils\cite{Guillamat2016b}. 
This suggests that submersed micropatterned structures can do more than impact flow and orientational state: we now demonstrate how substrate micropatterning can be used to control active matter concentration via structures raised above the solid substrate yet still fully submersed in the underlying oil-layer.

\subsection*{Pillars}

We consider fully submersed SU-8 pillar structures (\fig{fig:Set_Up}g) of radius $r_p=116\pm2\upmu\textrm{m}$ and height $h_p=6.8\pm0.3\upmu\textrm{m}$. 
As in the trench, sinusoid and stairway geometries, the active nematic layer is subject to a step-change in the effective film friction. 
However, differentiating it from previous structures, the pillar's virtual boundary forms a closed loop. 
The most prominent effect is a pronounced dilution of active material from the enclosed region above the pillar (\fig{fig:Set_Up}h; \mov{6}{mov:pillarExp}), which is qualitatively recapitulated in the simulations (\fig{fig:Set_Up}i; \mov{7}{mov:pillarSim}; \mthds{methods:sims}). 
The phase-field active material concentration $\phi\left(\vec{r};t\right)$ (\mthds{methods:sims}) demixes in the high effective friction region directly above the pillar. 
We confirm that the absence of active material is not a result of the pillar intruding through the active nematic film by occasionally observing microtubule bundles moving above the apex of the pillar (\mov{8}{mov:defectOverPillar}). 
Further, we did not observe curvature of the superjacent active layer in the vicinity of the micropillar, indicating that a finite oil depth lies between the subjacent pillar top and the nematic film. 

\begin{figure*}
    \centering
    \includegraphics[width=0.65\textwidth]{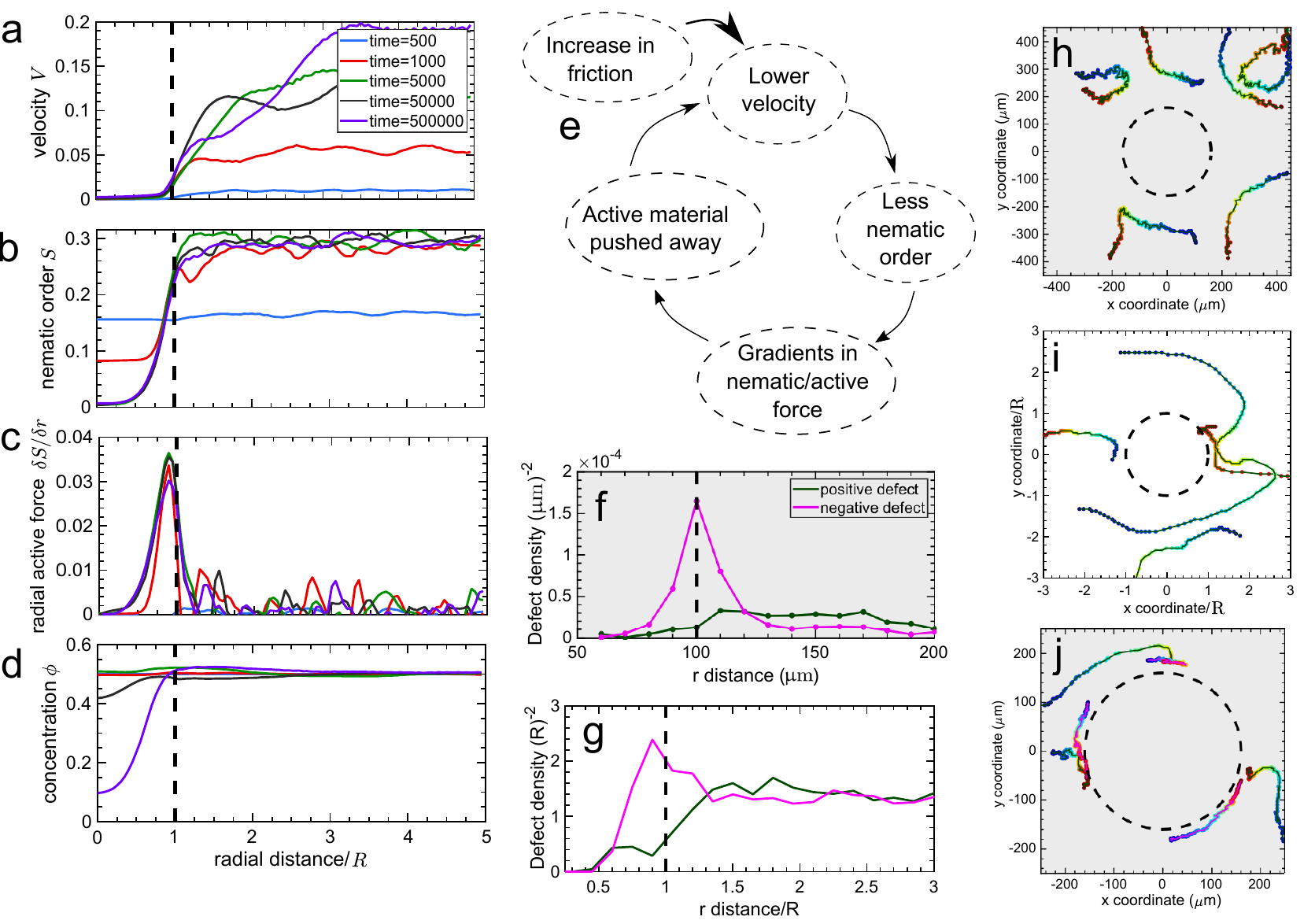}
    \caption{\small
	\textbf{Pillars cause local high friction regions which result in  active matter depletion. }
	(a) Due to the higher friction, the flow in the active nematic film remains low. 
	(b) The nematic is highly ordered far from the pillar but disordered above it because the speed is lower in the higher friction region. 
	(c) The difference in nematic order at the friction interface results in a radial active force.
	(d) This radial active force pushes the active material concentration outwards resulting in depletion effects. 
	Panels with non-shaded backgrounds denote numerical results, while shaded grey backgrounds denote experimental measurements. 
	(e) Schematic of a-d. 
	(f-g) $+1/2$ (dark green) and -1/2 (magenta) defect distribution as a function of radial distance from a submersed pillar from experiments (f) and simulations (g). 
	Dark green and magenta lines denote $+1/2$ and $-1/2$ defects, respectively. 
	(h-j) xy trajectories of example $\pm$1/2 defect dynamics in the vicinity of the pillar. 
	Time along the trajectory is displayed as circular markers coloured blue at the initial time and changing to red at the final instant. 
	(h-i) $+1/2$ defects deflecting from the pillar; experiments (h) and simulations (i). 
	(j) Experimental xy trajectories of $+1/2$ defects absorbing to the pillar boundary through defect annihilation with $-1/2$ residing in the vicinity of the perimeter. 
    }
    \label{fig:concentration_depletion}
\end{figure*}

To understand the mechanism leading to the pillar-bound dilute phase of active matter, we consider a simplified model of the active nematic. 
The effective friction is locally large above the pillar, causing the flow speed to decrease in the enclosed area but remain non-zero beyond the pillar border (\fig{fig:concentration_depletion}a).  
Since nematic ordering arises in active microtubule network films due to activity-induced motion, the sharp decrease in flow causes a corresponding drop in nematic scalar order $S$ across the circular virtual border (\fig{fig:concentration_depletion}b). 
However, the abrupt gradient in $S$ produces a radially outward average active force $\vec{f}(r;t) \sim \left\langle \bnabla\cdot\mathbf{Q}(\vec{r};t) \right\rangle \approx \partial_r \zeta(r;t)S(r;t)\unitvec{r}$ (\eq{eqn:activeStress} of \mthds{methods:sims}), when the variation of the nematic order is dominated by the radial change in scalar order parameter and bend-induced stresses are neglected. 
Thus, the active forcing is expected to be radially outward and sharply peaked about the interface as observed in simulations (\fig{fig:concentration_depletion}c). 
Interestingly, this argument suggests that contractile systems still deplete from the pillar region: While the $\zeta$ term will change sign, so will the $\partial_r S(r;t)\unitvec{r}$ since contractile flows lower the nematic order\cite{Thampi2015}. 
This is observed in simulations (\sifig{3}{sifig:contractile}). 

However, activity does not simply produce increased pressure across the perimeter but rather is able to selectively deplete the concentration of active material $\phi$, which enables depletion in incompressible films.
Since the active film is considered incompressible, fluid mass density is constant and hence divergence of the film velocity is zero (\eq{eq:u0} of \mthds{methods:sims}).
Thus, for depletion to occur, it is demanded that outward advection is more frequent in regions where $\phi$ is larger on average. 
This is indeed the case because the activity depends on the local amount of active material present, $\zeta(r;t)=\zeta_0\phi(r;t)$ (\eq{eqn:activity} of \mthds{methods:sims}), which causes the radially outward forcing to be stronger in magnitude where $\phi$ is large. 
For this reason, if the surrounding active turbulence stochastically advects $\phi$-rich active material across the perimeter, the local active forces increase in kind. 
Hence, the active forces selectively repulse the material from the dissipative region (\fig{fig:concentration_depletion}d; \mov{7}{mov:pillarSim}). 
However, the local active forces decrease when $\phi$-poor fluid enters. 
This allows the concentration to more easily cross the perimeter.
In this way, the depletion of active matter above the pillar is a result of the high effective friction lowering the velocity. 
This causes nematic discomposure and selectivity due to the direct dependence of activity on the local concentration (\fig{fig:concentration_depletion}e). 

Noting that the highest steps in the stairway geometry also fail to exhibit continuous nematic fields but are not devoid of active matter (\fig{fig:Set_Up}d; \mov{5}{mov:stairs}), we test if the curvature of the virtual barriers impacts depletion by simulating a rectangular pillar (\sifig{4}{sifig:rectangle}). 
We observe a comparable depletion of $\phi$ from the enclosed area as in the circular pillar and so conclude that curvature is not the critical difference. 
Secondly, we consider a circular pit where the friction is zero for radius $r<R$ and $\gamma=0.1$ for $r>R$. 
We find the active forces point inwards and observe an accumulation of $\phi$ (\sifig{5}{sifig:anti-pillar}). 
These inward active forces can even be used to trap two $+1/2$ defects for small radii, resulting in circulatory motion reminiscent of an active nematic in circular confinement (\sifig{6}{sifig:defect_trapping})\cite{Opathalage2019}. 

We conclude that accumulation or depletion of active material using submersed micropatterned structures relies principally on two attributes: 
(i) The oil-layer must be thin (high effective friction) to suppress the active flows necessary to exhibit nematic order. (ii) An enclosed area must be circumscribed by a virtual boundary to prohibit longitudinal active streams through the incompressibility constraint. 
Hence, we only find minute changes in $\phi$ to occur in the trench and stair geometries. 
Lastly, we point out that material does not deplete from the pillar centre in simulations on timescales investigated when a pillar radius is much larger, since the material does not diffuse to the interface where it could be selectively depleted. 

In addition to controlling concentration, submersed pillars interact with defects. 
We observe a greater frequency of $-1/2$ defects in the vicinity of the virtual boundary in our simulations (\fig{fig:concentration_depletion}f-g). 
The planar alignment of the director field explains the distribution of defects at the pillar boundary (\fig{fig:Set_Up}h-i). 
The resulting bend deformation around the perimeter drives hydrodynamic instabilities to continually generate defect pairs, with newly created self-motile $+1/2$ defects typically oriented radially away from the centre such that they swiftly move away from the interface (\mov{9}{mov:defectPillar}), leaving unbound immotile $-1/2$ defects behind 
\fig{fig:concentration_depletion}f-g). 

Submersed pillars can also serve as a virtual obstacle for defect trajectories (\mov{9}{mov:defectPillar}). 
Positive defects that approach the pillars from the surrounding turbulence stall or are deflected once in proximity to the pillar (\fig{fig:concentration_depletion}h-j). 
In Fig.~5h and j, we exhibit nine representative trajectories out of the 38 analyzed, and we show four from a total of 360 in \fig{fig:concentration_depletion}i to illustrate the observed interactions. 
Deflected $+1/2$ defects first slow as they approach the pillar, then scatter and regain speed as they move away from the submersed structure (\fig{fig:concentration_depletion}h-i). 
Positive defects that stall as they approach the boundary temporarily hold their position before annihilating with pillar-associated $-1/2$ defects (\fig{fig:concentration_depletion}j). 
While positive defects that directly approach the pillar can temporarily enter the depleted area by driving active material ahead of them, such infrequent events are transient (\mov{8}{mov:defectOverPillar}) as the repulsive active force (\fig{fig:concentration_depletion}c) pushes such incursions radially outward. 

\section*{Conclusions}

Using a combined experimental and simulation approach, we have demonstrated that micropatterns fully submersed in an underlying oil layer can guide the flow, topology, and even concentration of active material in superjacent nematic films without direct contact. 
By imposing changes in substrate depth, viscous dissipation in the oil layer enacts a position-dependent effective friction coefficient on the active material.
Abrupt substrate height steps can constitute sharp virtual boundaries in the active matter layer, which can control flow and defect behaviour. 
As proof-of-concept systems, we presented virtual channels exhibiting coexisting flow states, sinusoid substrates that gently guide defects, stairways separating active turbulence with differing characteristic length, and a virtual enclosure depleted of active material that acts as an obstacle scattering nearby defects. 

The proposed technique of fully submersing micropatterned structures facilitates new approaches for fabricating complex active topological microfluidic devices. 
For example, complications associated with infiltration of active nematics into confined spaces could be avoided, and active dynamics in various geometries at the same activity could be compared directly. 
Furthermore, locally concentrating or depleting active material could regulate activity at constant levels of ATP or rheological properties such as film viscosity or nematic elastic coefficient through this novel approach.

\section*{Materials and Methods}\label{sctn:methods}

\subsection*{Formation of the Active Nematic Network}\label{methods:nematic}

An active nematic microtubule network is generated following the protocol previously reported by Sanchez \textit{et al.} at an oil-water interface\cite{Sanchez2012}. 
Prior to experiments, active premixtures are prepared in $3.73\upmu\textrm{L}$ aliquots containing biotin-labeled K401 kinesin motors, streptavadin, Pyruvate Kinase/Lactic Dehydrogenase (PKLDH), phosphoenol pyruvate (PEP) (used for adenosine triphosphate (ATP) regeneration), $4\textrm{mg/ml}$ glucose, $0.27\textrm{mg/ml}$ glucose oxidase, $47\upmu\textrm{g/ml}$ catalase and $2\textrm{mM}$ Trolox, and $6\%$ (w/v) $20\textrm{kD}$ polyethylene glycol (PEG) in M2b buffer ($80\textrm{mM}$ piperazine-N,N'-bis(2-ethanesulfonic acid) (Pipes) $\textrm{pH} 6.8$, $2\textrm{mM}$ MgCl$_2$, and $1\textrm{mM}$ ethylene glycol-bis($\beta$-aminoethyl ether)-N,N,N',N'-tetraacetic acid (EGTA)). 
To prevent photobleaching during imaging, premixtures contain $6.65\textrm{nM}$ dithiothreitol (DTT), an antioxidant solution. 
Once the premix is prepared, the aliquots are flash frozen in liquid nitrogen and promptly stored at $-80^{\circ}\textrm{C}$ for future use. 

To perform active nematic experiments, $1\textrm{mM}$ ATP (final concentration) is added to the $3.73\upmu\textrm{L}$ premixture aliquot followed by $2\upmu\textrm{l}$ of $6\textrm{mg/ml}$ ($3\%$) Alexa Fluor 647 labeled guanylyl-(alpha, beta)-methylene-diphosphonate (GMPCPP) microtubules (for a final concentration of $1\textrm{mg/ml}$). 
For fluorescence imaging, microtubules are fluorescently labelled with Alexa Fluor 647\cite{Sanchez2012}. 
The result is a suspended active nematic with a total volume of $6\upmu\textrm{L}$. 
We use an ATP concentration at saturation, \textit{i.e.} the local microtubule extension speed was maximised. 
To confine the active nematic at an oil/water interface, we follow the previously published procedure\cite{Sanchez2012,Tan2019}. 
We first create a flow cell made from the glass substrate with patterned structures, double-sided tape and a coverslip treated with a polyacrylamide brush (\fig{fig:Set_Up}). 
The polyacrylamide brush prevents excess protein binding to the coverslip. 

We flow in an oil/surfactant mixture (Novec HFE 7500 Engineered Fluid with $1.8\%$ PFPE-PEG-PFPE (perfluoropolyether) surfactant) into the channel. 
Then, this mixture is exchanged with the active microtubule network. 
We ensure that the amount of oil in the channel is constant over time after filling by sealing the flow cell after preparation. For filling consistency, each cell is prepared using identical spacers and volumes of oil; however, some variation in oil thickness is expected due to microfabrication tolerances. 
This system forms a 3D, unconfined active microtubule network. 
Streptavidin can bind up to four biotin-labeled kinesin molecules and when microtubules of opposing polarities align parallel to each other, the kinesin molecules oriented at $180^{\circ}$ to each other walk in opposite directions along those neighbouring microtubules. 
As the kinesins walk, the filaments produce an extensile motion driven by ATP hydrolysis. 
The ends of the flow cell are sealed using a UV-curable glue (RapidFix). 
The active layer is then centrifuged using a swinging bucket rotor for 42 min at $300\textrm{rpm}$. 
The dynamic viscosity of the oil layer is $1.24\times10^{-3}\textrm{Pa}\cdot\textrm{s}$ and the film viscosity of the active nematic gel has previously been found to be $\sim10^{-3}\textrm{Pa}\cdot\textrm{s}\cdot\textrm{m}$\cite{Guillamat2016b}. 

\subsection*{Photolithography}\label{methods:photo}

The trench and pillar geometries are produced using photolithography. 
SU-8 (MicroChem Corp.) is a negative tone epoxy-based photoresist that is used to create thin film plastics on substrates. 
SU-8 is composed of epoxy-based monomers and photo-acid generators (PAGs) suspended in a solvent. 
Upon exposure to ultra-violet (UV) light, the PAGs release acids serving as a catalyst for cross-linking available epoxy groups on the monomer once heat is applied to the substrate. 
Prior to fabrication, glass substrates are thoroughly cleaned in soap and water followed by 30 minutes of sonication in acetone, methanol and ethanol in this order. 
The glass substrates are rinsed in nanopure water to ensure minimal presence of surface contaminants. 
The glass is then plasma treated with oxygen for 2 minutes. 
To ensure the removal of residual moisture left on the surface, the glass substrate is placed on a hotplate for 5 minutes at $200^{\circ}\textrm{C}$; the substrate is left to cool down to room temperature for 5 minutes in a humidity-controlled environment. 

Upon thorough cleaning, a quarter-sized drop of SU-8 50 is deposited on the glass substrate. 
The SU-8 is spin-coated at $2000\textrm{rpm}$ for 45 seconds followed by a 10 minute wait. 
The substrate is then soft baked at $65^{\circ}\textrm{C}$ for 12 minutes then at $95^{\circ}\textrm{C}$ for 45 minutes on a hotplate to evaporate solvent. 
The substrate is again left to cool to room temperature with another wait of 10 minutes. 
The film is then exposed to $365\textrm{nm}$ UV light ($500 \textrm{mJ/cm}^{-2}$) followed by a 10-minute wait step. 
To crosslink the exposed regions' epoxy groups, the substrate undergoes a post exposure bake for 5 minutes at $65^{\circ}\textrm{C}$ then for 15 minutes at $95^{\circ}\textrm{C}$ on a hot plate. 
After the 15 minute bake, the hot plate is turned off and allowed to cool to room temperature without the removal of the coated substrate. 
This step is to avoid thermally shocking the film which can result in cracks and poor adhesion. 
The substrate is developed for 30 minutes with gentle agitations. 
Once developed, the residual SU-8 developer is rinsed with isopropanol and de-ionised water then dried with nitrogen gas. 
After development, the substrate is then hard baked for 2 minutes at $150^{\circ}\textrm{C}$. 
Internal stresses within the substrate due to the abrupt change in surface height are apparent near the edge of the trench as parallel lines in \fig{fig:Set_Up}b. This accounts for the gap in data at the trench edge in \fig{fig:Defect_depletion}c. 
We note that some local residual SU-8 is present at the base of the pillar in \fig{fig:Set_Up}h, but this material does not appear to affect the active nematic flow dynamics. 
The heights of the microstructures are measured using profilometry. 

\subsection*{Micro-milling}\label{methods:milling}

The stepped substrate and the sinusoid surface are produced using the computer numerical controlled (CNC) micromilling. 
We use the milling machine TN5-V8-TC8 (MDA Precision) to fabricate the microstructures from poly(methyl methacrylate), commonly known as acrylic. 
This milling machine is capable of handling drills and endmills as small as $50\upmu\textrm{m}$ in diameter and has a spindle rotational accuracy of around $2\upmu\textrm{m}$. 
The first designed system has steps (height), each with a rise of $\Delta_s = 10 \pm 1\upmu\textrm{m}$ and a run (horizontal width) of $w_s = 500 \pm 2 \upmu\textrm{m}$. 
The system is composed of ten steps. 
The smaller of the two sinusoid systems has undulations with a peak-to-peak amplitude $\Delta_u = 40 \pm 2 \upmu\textrm{m}$ and a wavelength $\lambda_u=150 \pm 2 \upmu\textrm{m}$, while the larger has  a peak-to-peak amplitude $\Delta_u = 50 \pm 2 \upmu\textrm{m}$ and a wavelength $\lambda_u=150 \pm 2 \upmu\textrm{m}$. 
After milling, these features are used as a mold and transferred into a soft elastomer, polydimethylsiloxane (PDMS), which in turn is used as a mold again to transfer the features into a cyclic olefin copolymer (COC), using a thermopress. 
The COC sheets of thickness ranging from $150 \upmu\textrm{m}$ to $350 \upmu\textrm{m}$ are then immersed for one hour in a solution of 10wt\% of 8-Anilino-1-naphthalenesulfonic acid (ANS), a hydrophobic dye, dissolved in a mixture of 85wt\% ethanol and 15wt\% decalin. 
This treatment adds a fluorescent layer of ANS dye to the COC surface, which allows us to use a 944 Leica SP8 UV/Visible Laser Confocal Microscope to measure the fluorinated oil’s thickness under the active nematic.
The confined active nematic is imaged using a wide-field fluorescence Nikon Eclipse Ti-E microscope with an Andor Clara camera controlled by \textit{Micromanager} open-source software. 

\subsection*{Data processing}\label{methods:data}

To investigate how defect dynamics in the active layer are influenced by submersed structures, labeled microtubule bundles are imaged using fluorescence microscopy. 
Four-hundred-frame videos are collected at 1 frame per second and processed using Fiji/ImageJ version 1.52a software. 
To acquire defect distributions, active nematic microtubule defects are identified and counted manually every 10 frames for each video. 
Two-dimensional Cartesian components for both x and y axes are acquired from both +1/2 and -1/2 defects using the \textit{Click-Coordinates-Tool} on Fiji/ImageJ. 
For the submersed trench geometry, we use MATLAB to analyse the frequency and position for both $+1/2$ and $-1/2$ defects across the channel. 
Defects are organised and binned in $10\upmu\textrm{m}$ horizontal increments across the field of view. 

For the stairway geometry and pillar geometries, we apply the same counting procedure to obtain $+1/2$ and $-1/2$ defect positional frequencies across all frames. 
In the stairway geometry, we apply the counting procedure sequentially to each step and, for the pillar, the videos are processed by centring the pillar in a $400\times400\upmu\textrm{m}^2$ window. 
Similar to the analysis done with the submersed trench, we use MATLAB to generate a 2D histogram to represent the frequency of the $+1/2$ and $-1/2$ defects, positionally distributed in a 2D plane. 
Defects are organised and binned every $13.4\upmu\textrm{m}$ in both horizontal and vertical increments. 

The nematic director field inside the channel was calculated using a Fourier transform-based method reported recently by our group\cite{Tan2019}. 
To measure the director orientation directly above trenches, the imaged active nematic is oriented with the long axis parallel to the y-axis. 
Each pixel is converted to micrometers ($0.9434\upmu\textrm{m}$/pixel). 
Each pixel from each frame containing a x-component, y-component and angle of the bundle-microtubule director is represented on a 2D grid and determined using MATLAB. 
Our process uses a nested loop and a conditional statement to determine if the angle is between $80$ and $90^\circ$; if a director satisfied this condition along the y-axis for the specified x-position, the total is summed then divided by the total number of angles checked by the loop. 
This probability is appended to a new horizontal array for each probability in the x-position. 
The result is a time-averaged director orientation mapped across the trench, averaging over all y values. 

\subsection*{Simulations}\label{methods:sims}
To complement the experiments, we simulate the active nematic thin film using a 2D hybrid lattice Boltzmann/finite difference approach. 
The incompressible active nematic film flows with velocity $\vec{u}\left(\vec{r};t\right)$, has long-range orientational order described by the tensor order parameter $\tens{Q}\left(\vec{r};t\right)$ and varying concentrations of active material, which we take to be a phase field $\phi\left(\vec{r};t\right)$ varying from 0 to 1, coarsely describing the local amount of active materials (microtubules, kinesin complexes and ATP). 
The total free energy includes nematic bulk (LdG) and deformation (FO) terms, as well as a binary mixture bulk (DW) and an interfacial (I) term; $\mathcal{F}\left[\tens{Q},\phi\right]=\int d^2\vec{r} \left( f_\text{LdG} + f_\text{FO} + f_\text{DW} + f_\text{I} \right)$. 
Four coupled equations describe the time evolution of the continuous fields. 

The first is the Beris-Edwards equation for nematics,
\begin{eqnarray}
  \label{eq:Q}
  \left(\partial_t + \vec{u} \cdot \bnabla\right)\vec{Q} - \vec{S} & = & \Gamma_Q \vec{H} . 
\end{eqnarray}
The co-rotation term $\vec{S} = \left(\xi \vec{D}+\vec{\Omega}\right)\left(\vec{Q}+\frac{1}{3}\vec{I}\right) + \left(\vec{Q}+\frac{1}{3}\vec{I}\right)\left(\xi \vec{D}-\vec{\Omega}\right)-2\xi\left(\vec{Q}+\frac{1}{3}\vec{I}\right)\text{tr}\left(\vec{Q}\vec{W}\right)$ determines the alignment of the microtubules in response to gradients in the velocity field, with $\tens{\Omega}$ the rotational part and $\tens{D}$ is the extensional part of the velocity gradient tensor $\tens{W}=\bnabla \vec{v} =
\tens{\Omega} + \tens{D}$. 
The alignment parameter $\xi$ is taken to be in the flow aligning regime and set to $\xi=0.5$. 
The resulting Leslie angle\cite{thijssen2020c,thijssen2020b} is $\theta_\text{L}=\tfrac{1}{2}\cos^{-1}\left(\tfrac{\xi(3S+4)}{9S}\right)$. 

The  molecular field $\tens{H} = -(\frac{\delta\mathcal{F}}{\delta\tens{Q}} - \frac{1}{3}\tens{I} \; \text{Tr}\frac{\delta\mathcal{F}}{\delta\tens{Q}})$ is a functional derivative of free energy density $\mathcal{F}$, describing the relaxation towards equilibrium at a rate $\Gamma_Q$. 
The free energy depends directly on the nematic tensor $\tens{Q}$ and the active material concentration $\phi$. 
The nematic part consists of a Landau-De~Gennes contribution, $f_\text{LdG} = A_0\left\{\tfrac{1}{2}\left(1-\tfrac{\nu}{3}\right)\text{tr}\left[\tens{Q}^2\right] - \tfrac{\nu}{3}\text{tr}\left[\tens{Q}^3\right] + \tfrac{\nu}{4}\text{tr}\left[\tens{Q}^2\right]^2\right\}$ with $A_0=0.05$, and Frank-Oseen deformation $f_\text{FO} = \tfrac{K}{2}\left( \bnabla \tens{Q}\right)^2$ with $K=0.02$. 
We set $\nu=2.55$, which favours the isotropic state in the absence of active flows\cite{Thampi2015} and is independent of $\phi\left(\vec{r};t\right)$. 
This choice allows nematic ordering only due to activity-induced flows, in agreement with experiments\cite{Sanchez2012} and previous simulations\cite{Thampi2015,Srivastava2016}.  Throughout the simulations, the nematic order $S$ is found to be between $0$ and $0.3$. 
For the trench data of \fig{fig:dif_flows}c, we find $S=0.18$ and calculate a Leslie angle of $\theta_\text{L}=22^\circ$. 

The active material concentration evolves according to a Cahn-Hillard model.
We assume that the film is incompressible and active material concentration $\phi$ does not impact fluid mass density $\rho$. 
\begin{eqnarray}
  \partial_t \phi + \bnabla \cdot \left(\vec{u}\phi\right) & = & \Gamma_\phi \nabla^2 \mu, \label{eq:phi}
\end{eqnarray}
where $\mu = \tfrac{\delta\mathcal{F}}{\delta\phi} - \bnabla \cdot\left(\tfrac{\delta\mathcal{F}}{\delta\bnabla\phi}  \right)$ is the chemical potential and $\Gamma_\phi=0.1$ is a mobility coefficient.  
The double-well free energy density depends on phase as $f_\text{DW}=\tfrac{A_\phi}{2}\phi^2\left(\phi-1\right)^2$ and an interfacial term $f_\text{I} = \tfrac{ K_\phi}{2}\left(\bnabla\phi\right)^2$ with $A_\phi=0.03$ and $K_\phi=0.1$.
The free energy minima are at $\phi=\{0,1\}$. 
While positive $A_\phi$ favours phase separation, activity suppresses demixing and $\phi$ does not phase separate into high and low $\phi$ regions for sufficiently active flows. 
In our model, active flows above high friction regions (such as within the perimeter of the pillar geometry) can become sufficiently small for spinodal decomposition to occur. 
This does not happen in the low friction region. 
We have verified that depletion in the pillar geometry also occurs for a binary mixture free energy modelled using a single-well, which would not thermodynamically phase separate. 
We initialise the concentration to $\phi=0.5$. 
All reported phenomena are due to dynamical interactions, since the free energy does not favour global nematic ordering nor phase separation. 

Lastly, the system obeys the Navier-Stokes equations for the velocity field within the active film. 
Assuming constant fluid mass density $\rho$ (not active material concentration $\phi$) leads to the incompressibility condition
\begin{eqnarray}
  \bnabla\cdot\vec{u} & = & 0. \label{eq:u0} 
\end{eqnarray}
Because the experimental active nematic film lies on a 2D interface between two 3D fluids, the planar divergence of the velocity could conceivably be non-zero. 
However, non-zero divergent velocity fields would require continuous circulation in the thin aqueous/oil layers above/below the film. 
As we are unaware of evidence indicating such flows, we assume 2D incompressibility implying divergence-free flow. 
Since the fluid is taken to be incompressible, any outward fluid mass flux into an enclosed area must be balanced by an inward flux and vice versa. If this constraint is relaxed, we allow flow sources in our film and non-selective outward pointing forces $\zeta\partial_r S(r;t)\unitvec{r}$, where the $\zeta$ is independent of $\phi$, will also deplete the material. 

The second equation is the Cauchy momentum equation
\begin{eqnarray}
  \rho\left(\partial_t + \vec{u} \cdot \bnabla\right)\vec{u} & = & -\bnabla p + \bnabla \cdot \vec{\Pi} - \gamma \vec{u}, \label{eq:momentum}
\end{eqnarray}
where $p$ is the pressure and $\tens{\Pi}$ is the stress tensor which includes the standard viscous stress $\tens{\Pi}^\textmd{visc} = 2 \eta  \tens{E}$ for film viscosity $\eta=2/3$. 
Furthermore, it contains the elastic stress due to the nematic nature of the microtubules
\begin{align}
\tens{\Pi}^{\text{elastic}} &= 2 \xi \tens{\mathcal{Q}} (\tens{Q}:\tens{H}) - \xi \tens{H}\cdot\tens{\mathcal{Q}}  - \xi \tens{\mathcal{Q}} \cdot \tens{H} \nonumber\\
                            &\quad -\bnabla\tens{Q} : \frac{\delta \mathcal{F}}{\delta \bnabla\tens{Q}} + \tens{Q}\cdot\tens{H} - \tens{H}\cdot\tens{Q},
\end{align}
where $\tens{\mathcal{Q}}=\tens{Q}+\tens{I}/3$. 
The stress also contains the capilary stresses $\tens{\Pi^{\text{cap}}}= (\mathcal{F}-\mu\phi)\tens{I}-\bnabla\phi\left(\frac{\delta\mathcal{F}}{\delta\bnabla\phi}\right)$ due to differences in concentration $\phi$, and the active component 
\begin{align}
    \tens{\Pi^{\text{act}}} &= -\zeta\left(\vec{r};t\right) \tens{Q} = -\zeta_0 \phi\left(\vec{r};t\right) \tens{Q}. 
                            \label{eqn:activeStress}
\end{align}
Here, we have taken the local activity 
\begin{align}
  \zeta\left(\vec{r};t\right)=\zeta_0 \phi\left(\vec{r};t\right) 
                            \label{eqn:activity}
\end{align}
to scale directly with active material concentration. 
We set $\zeta_0=0.09$.
While the velocity and director fields can in principle develop out-of-plane components in the nematic film, in experiments and for the parameter values used in this study, no out-of-plane components develop\cite{Doostmohammadi2017}. Comparison to strictly two-dimensional simulations has confirmed the qualitative agreement\cite{hemingway2016correlation}. 

In the Navier-Stokes equation, we also include an effective friction $\gamma\left(\vec{r}\right)$. 
Employing an effective friction that varies with oil-layer depth neglects any recirculation effects within the oil layer, which is justified by the facts that the micropatterned structures are characterized by small height/width ratios and the Reynolds number of flows within the oil layers are infinitesimal. 
Hence, typical recirculation vortices around pillars are neglectable\cite{baykal2015numerical}. 
Since previous studies have successfully modelled the dynamics of microtubule/kinesin-based active nematic films with weak effective friction\cite{Zhou2020}, we treat the friction as negligible in the regions of the film superjacent to deep structures. 
In the regions above shallows, we include lubrication momentum dissipation via a non-zero effective friction coefficient. 
We set $\gamma(h_s)=0.07$ in the shallows for simulations of the trench system. 
We define a characteristic confinement scale to be $R\equiv20$ lattice Boltzmann (LB) nodes. 
This is comparable to the intrinsic active nematic length scale and, as a result, the reported defect densities are of order unity. 
The narrowest trench width is $w_t=1.2R$ (\fig{fig:dif_flows}c) and the two wider trenches have widths of f $w_t=1.7R$ (\fig{fig:dif_flows}b) and $6R$ (\fig{fig:dif_flows}d), respectively. 
All trenches are simulated in $7.5R\times30R$ periodic systems. 
For the stairway, we simulate a $7.5R\times50R$ long system composed of the same number of steps as in the experimental system (10 steps) of width $w_s=5R$. 
Each step represents a different lubrication momentum dissipation region with different non-zero effective friction coefficients set to $\gamma(h_s,n)=\frac{0.1}{1+n}$, where $n\in \{0\ldots9\}$ denotes the different steps corresponding to different oil depths in the experiments. 
In the main text, we present simulation results from steps $n=\{1,2,3,4,5\}$. 
For the pillars, we use a radius $r_p=R$. 
The friction coefficient is chosen to be sufficiently large to clearly match the pronounced depletion observed experimentally in \fig{fig:concentration_depletion}. 
We present results for the value $\gamma(h_p)=0.5$ but observe depletion of $\phi$ for $\gamma(h_s)=0.07$, corresponding to the choice used for the trench geometry. 
The pillar is simulated in a $10R\times10R$ periodic system. 

Defects are detected by calculating the winding number\cite{hobdell1997numerical}. If the average nematic order $S$ in a range of 3 LB nodes around a defect is lower then $0.05$, we assign this defect to be virtual and ignore it.

\section*{References}

\bibliography{refMicropatterned}


\section*{Acknowledgements}
We thank Amin Doostmohammadi for helpful discussions. We acknowledge generous funding from the National Science Foundation, through several awards. 
LSH and DAK thank DMR-1808926 and NSF-CREST: Center for Cellular and Biomolecular Machines at UC Merced (HRD-1547848), LSH, DAK, SF and MG are grateful to the Brandeis Biomaterials facility MRSEC-2011486. 
K.T. acknowledges funding from the European Union's Horizon 2020 research and innovation programme under the Marie Sklodowska-Curie grant agreement no. 722497 (LubISS) and the EPSRC grant EP/T031247/1. 
T.N.S received funding from the European Union's Horizon 2020 research and innovation programme (grant agreement no. 851196).


\section*{Supplementary Information}\label{sctn:si}


\setcounter{figure}{0}
\renewcommand{\figurename}{FIG. S\hspace{-0.3em}}

\subsection*{Sinusoidal Microstructures}
To demonstrate that the method of submersed micropatterned substrates in the oil layer below an active nematic film can guide defect dynamics, we produced a series of micropatterned sinusoidal substrates (\mthds{methods:milling}). 
The smallest of these are characterised by amplitude $\Delta_u = 40 \pm 2 \upmu\textrm{m}$ and wavelength $\lambda_u=150 \pm 2 \upmu\textrm{m}$. 
As observed in \mov{mov:undulSmall} and discussed in the main text, the undulating effective friction produces orientation-control of motile defects for smaller wavelengths. 
Trains of co-aligned +1/2 defects move along the troughs (\sifig{1}{sifig:sinusoid}a). 
However, these are intermittently disrupted by the extensile-active nematic hydrodynamic-bend instability. 
Regions of well-ordered nematic aligned along the troughs result in pair creation events in which +1/2 defects unbind with an orientation perpendicular to the troughs (\sifig{1}{sifig:sinusoid}b). 
\begin{figure}
    \centering
    \includegraphics[width=0.45\textwidth]{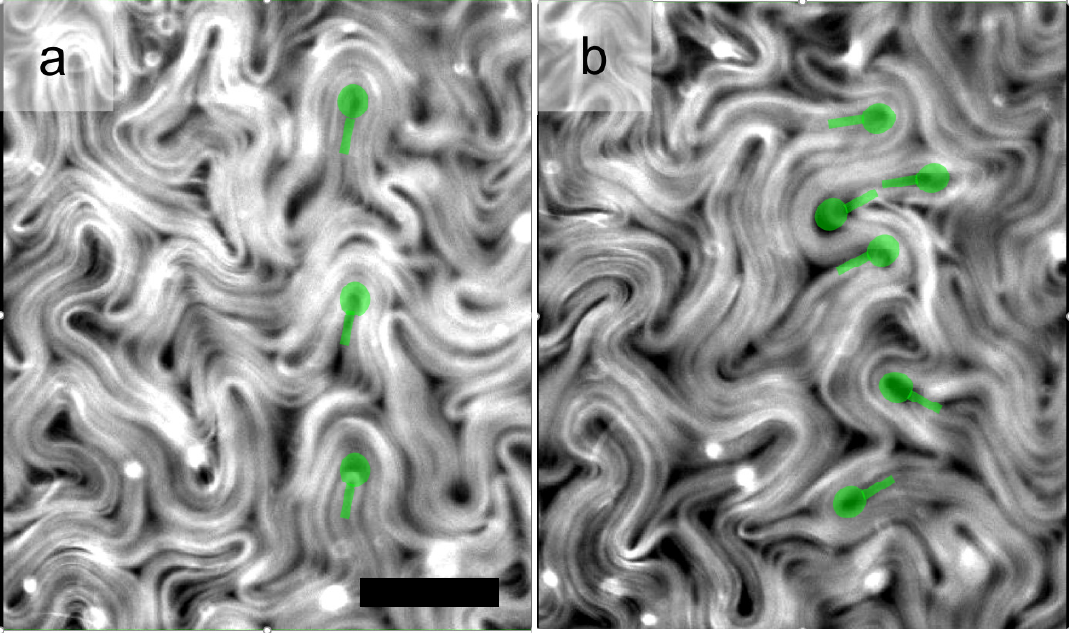}
    \caption{\small
	\textbf{Submersed micropatterned sinusoidal substrate.} 
	Fluorescent image of the active nematic film superjacent to an oil-submersed sinusoidal micropatterned structure with a $150\upmu\textrm{m}$ wavelength and $40\upmu\textrm{m}$ amplitude. 
	The undulations go from left to right such that the parallel troughs and crests run in parallel up and down. 
	Scale bar $= 250 \upmu\textrm{m}$.
	(a) A set of three motile +1/2 defects that move in a train up the middle of a trough are highlighted in green. 
	The train aligns the director parallel to the trench. 
	(b) $15$ seconds later the initial train is destabilised by extensile-active nematic hydrodynamic-bend instability and the associated pair creation events in which +1/2 defects tend to unbind oriented perpendicular to the trough. 
    }
    \label{sifig:sinusoid}
\end{figure}

\subsection*{Confocal Microscopy Measurement of the Oil Thickness}
We report on the positional defect density above submerged acrylic micro-milled steps. 
Consistent oil layer thickness is controlled indirectly through the height of the spacers and the oil's filling volume. 
The initial thickness of fluorinated oil layer in the stairway geometry is $h_{0} \approx 12 \pm 3 \upmu\textrm{m}$ followed by $10 \upmu\textrm{m}$ deep increments along the x-direction. 
Confocal microscopy provides a 3D reconstruction of the active film to demonstrate that the active layer is without deformation. This verifies that the active layer itself is consistently flat, even above micropatterned structures. 
The solid substrate supporting the fluorinated oil is also observed to be smooth and flat in the absence of micropatterns. The difference between the positions of the oil interface and solid substrate, together with the refractive index of the oil, allows us to estimate the thickness of oil. 
 \begin{figure}
          \centering
          \includegraphics[width = .95\linewidth]{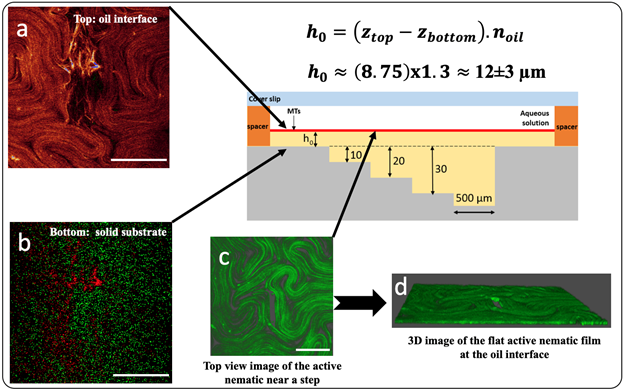}
          \caption{
          \textbf{Confocal microscopy measurement of the fluorinated oil’s thickness.} (a) Confocal fluorescent images of the active nematic film, (b) the solid substrate supporting the fluorinated oil, and (c) the active nematic film near a step. The thickness of oil $h_{0} \approx 12 \pm 3 \upmu\textrm{m}$ is obtained by measuring the difference between the heights of the active nematic film and the solid substrate $(z_{top} - z_{bottom}) \approx 8.75 \upmu \textrm{m}$, multiplied by the refractive index of the oil $n_{oil} \approx 1.3$.  (d) Near the steps, the 3D reconstruction of the active nematic confirms the film is flat without any deformation. Scale bars $= 100 \upmu \textrm{m}$.} 
     \label{sifig:SIconfocalmeasurement}
\end{figure}
         
\subsection*{Depletion in Contractile Active Nematic Systems}
In the main text, we explain the selective depletion of active material $\phi$ from the region superjacent to the pillar by considering the active force $\vec{f}(r;t) \sim \left\langle \bnabla\cdot\mathbf{Q}(\vec{r};t) \right\rangle \approx \partial_r \zeta(r;t)S(r;t)\unitvec{r}$ over the radial interface. 
For the extensile active nematic bundled-microtubule films considered in the text, the positive activity $\zeta$ and actively ordered surroundings, produce a radially outward force. In contractile systems ($\zeta<0$), it has been established that the nematic order $S$ will decrease due to active flows. Hence, both factors change sign and our argument still predicts depletion for such contractile systems because the product does not change sign. We verify this argument numerically (\sifig{3}{sifig:sinusoid}) in simulations with activity $\zeta=-0.03$ and free energy parameter $\nu=3$, which results in nematic order in the absence of activity. We find that the nematic order $S$ decreases across the pillar perimeter (\sifig{3}{sifig:sinusoid}; red line) and that depletion from the region above the pillar still occurs (\sifig{3}{sifig:sinusoid}; purple line). 
\begin{figure}
  \centering
  \includegraphics[width=0.45\textwidth]{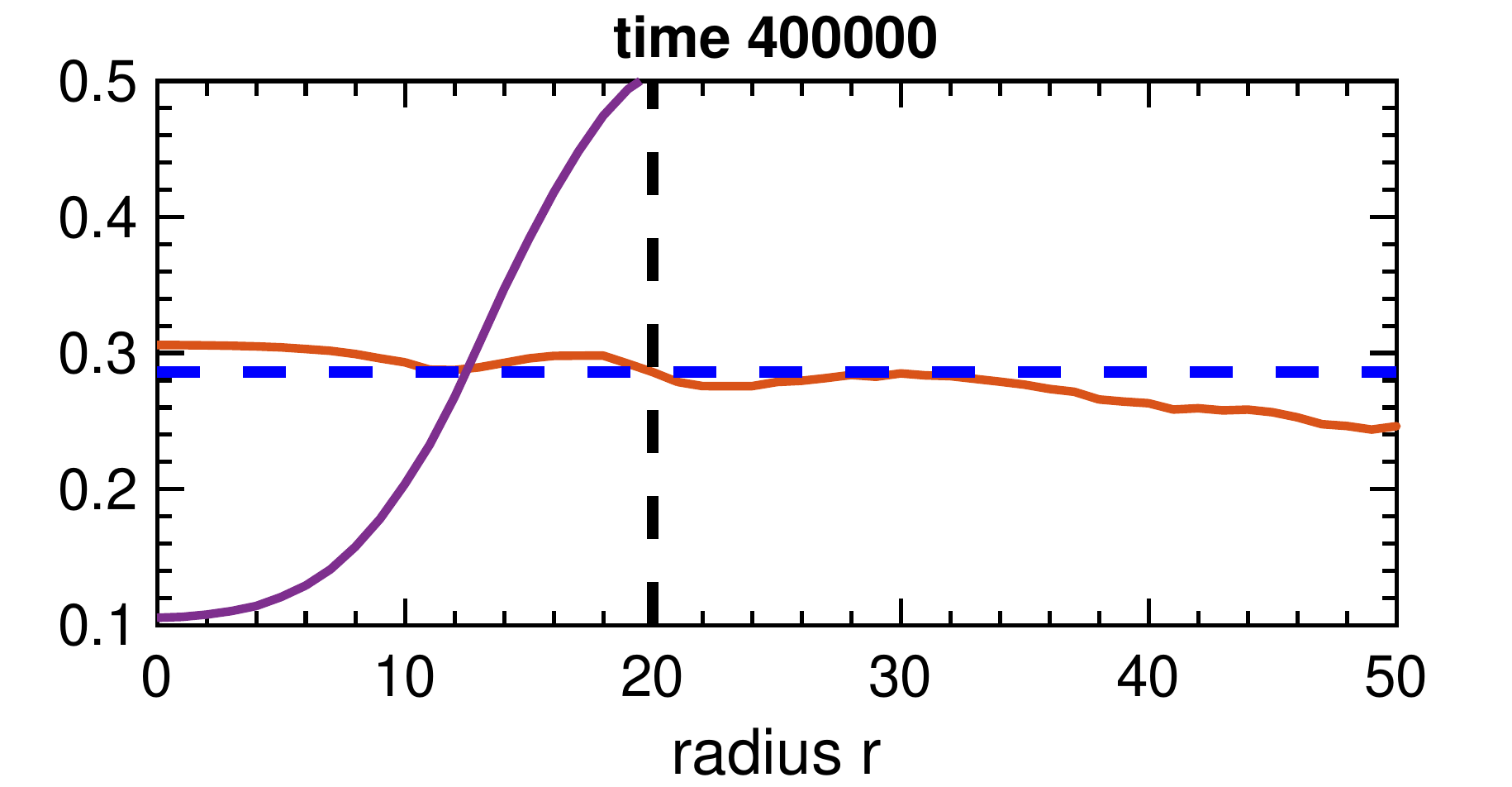}
  \caption{\small
    \textbf{Contractile stresses still result in depletion.} 
    A contractile system ($\zeta=-0.03$) with a free energy parameter $\nu=3$. The friction is $\gamma=0.5$ for $r<20$ and zero outside. The purple line is the active nematic concentration $\phi$ and the red line is the nematic order $S$. With this value of free energy parameter $\nu$, the equilibrium  nematic ordering $S_{eq}\neq0$ and decreases due to contractile flows outside the pillar. The blue dashed line is a visual guide to help the reader identify the change in $S$.} 
  \label{sifig:contractile}
\end{figure}

\subsection*{Rectangular Microstructure Pillar}
Simulations of rectangular pillar (\sifig{4}{sifig:rectangle}) exhibit a comparable depletion of concentration $\phi$ from the enclosed area above the pillar as the circular pillars considered in the main text (\fig{fig:concentration_depletion}). 
Hence, we conclude that a pillar curvature is not the critical property leading to depletion in the pillar geometry. 
Rather, the abrupt gradient in $S$, which follows indirectly from the increased effective friction, produces an outward average active force that can deplete active material from any enclosed area of high effective friction. 
\begin{figure}
  \centering
  \includegraphics[width=0.45\textwidth]{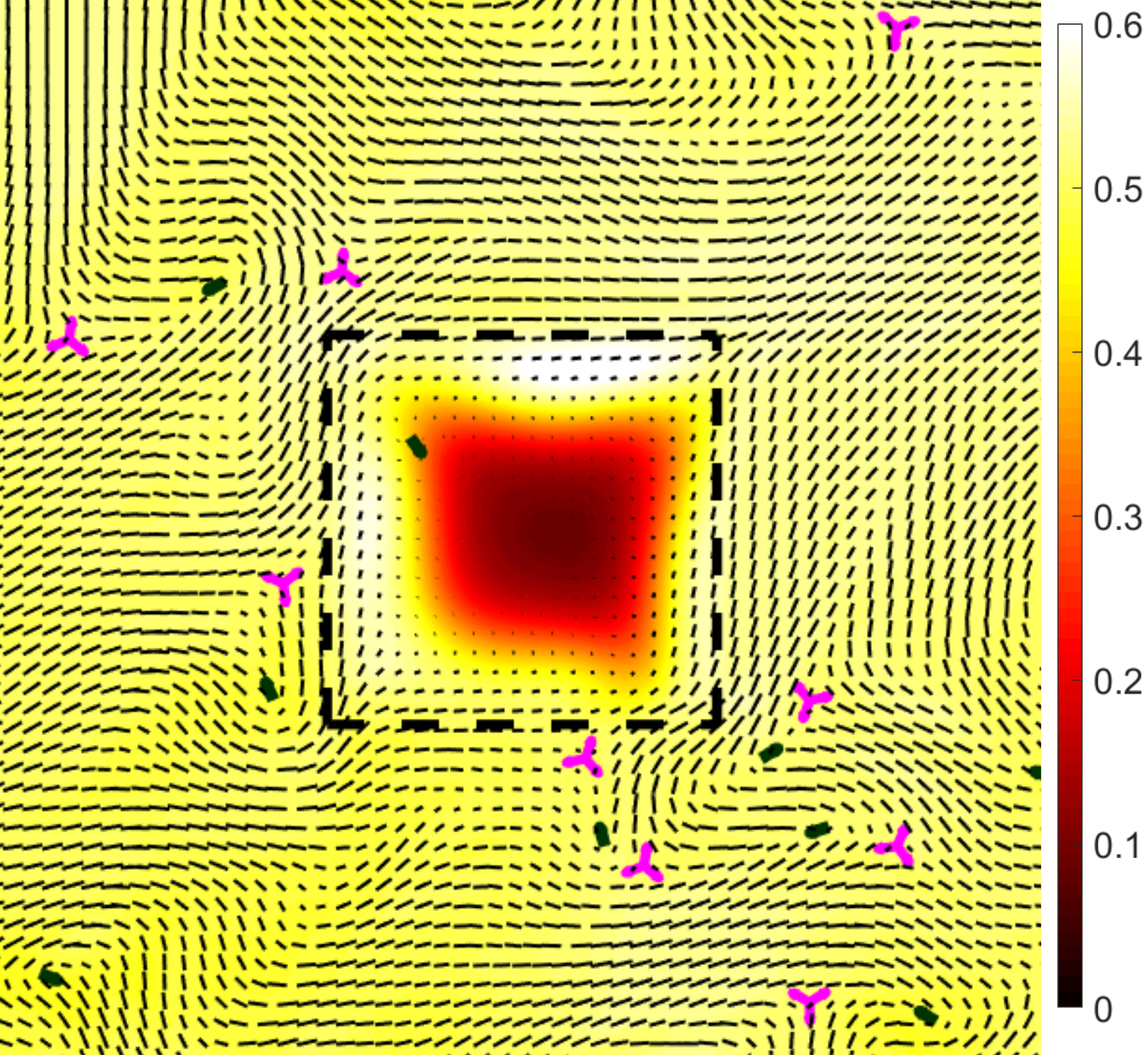}
  \caption{\small
    \textbf{Rectangular pillar causes depletion.} 
    Same as \fig{fig:Set_Up}i but with the circular pillar replaced with a square, the perimeter of which is traced by dashed lines. 
    We observe a comparable degree of depletion of active material concentration $\phi$, denoted by the colormap, for the same friction coefficient $\gamma=0.5$. 
  }
  \label{sifig:rectangle}
\end{figure}

\subsection*{Microstructure Pit}
In the main text, we argue that active material is depleted from the enclosed region directly above a microstructured pillar due to the increased effective viscous dissipation within the locally thinner oil-layer. 
Thus, one naturally wonders if a microstructured pit in the substrate might act as an ``anti-pillar'' and lead to accumulation of active material within an enclosed region above the structure. 
We consider an effective circular pit by simulating a substrate with an effective friction except directly above the structure (\sifig{5}{sifig:anti-pillar}) and observe that active material accumulates in the enclosed area. 
The velocity magnitude, nematic scalar order parameter, and concentration directly above the pit (\sifig{5}{sifig:anti-pillar}) are all comparable to their values far from the pillar (\fig{fig:concentration_depletion},a,b,d respectively). 
While the change in the scalar order parameter is positively peaked at the perimeter of the pillar (\fig{fig:concentration_depletion}c), it is seen to dip at the perimeter of the pit (\sifig{5}{sifig:anti-pillar}). 
This indicates a radially inward average active force, consistent with the mechanism which here leads to accumulation. 
\begin{figure}
  \centering
  \includegraphics[width=0.45\textwidth]{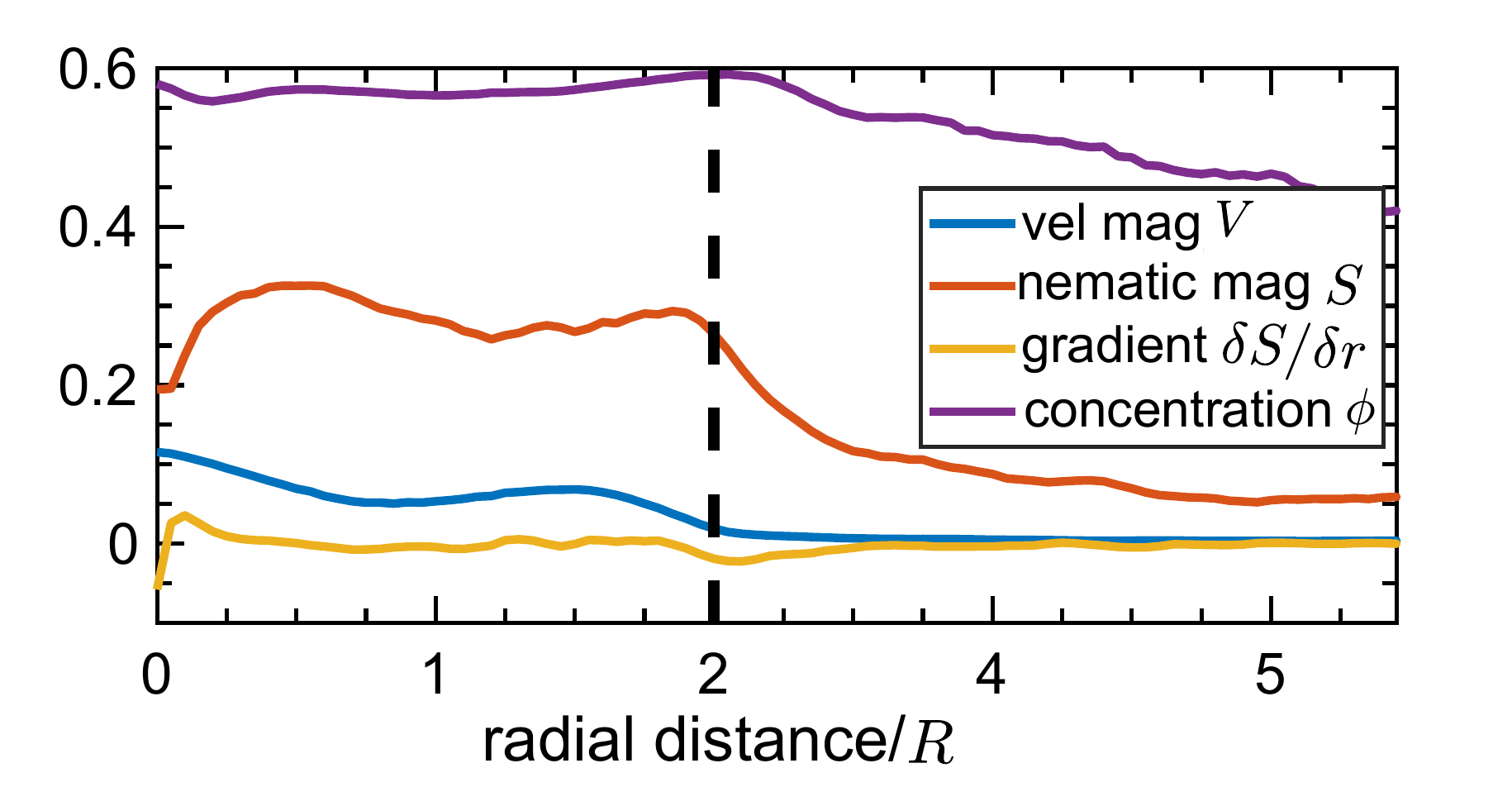}
  \caption{\small
    \textbf{Pits cause local lower friction regions which result in active matter accumulation.} 
    We consider a circular pit of radius $2R$ above which the frictional damping is negligible but outside of which $\gamma=0.1$. 
    This friction term is lower than for the pillar geometry presented in the manuscript. This is necessary to ensure spinodal decomposition does not occur outside the geometry.  
    We measure the same profiles as \fig{fig:concentration_depletion}a-d for one instant after steady state has been achieved. 
    We observe accumulation of active material $\phi$ in the enclosed area. 
  }
  \label{sifig:anti-pillar}
\end{figure}

\subsection*{Microstructure Pits Can Trap Defects}
Depletion occurs due to radially outwards active forces, such that the region superjacent to micropatterned pillars can act as obstacles for approaching defects.
On the other hand, the pit geometry produces a radially inward average active force (\sifig{5}{sifig:anti-pillar}). Together, these two facts suggest that pits may trap defects within their perimeters. By varying the pit radius such that it is comparable to the intrinsic active nematic length scale in simulations, we indeed find a regime where two $+1/2$ defects can be trapped in the interior of the pit for extended periods and rotate around each other (\sifig{6}{sifig:defect_trapping}). Similar behaviour has been observed for pairs of defects trapped by impermeable circular confining walls. 
\begin{figure}
  \centering
	\includegraphics[width = .45\textwidth]{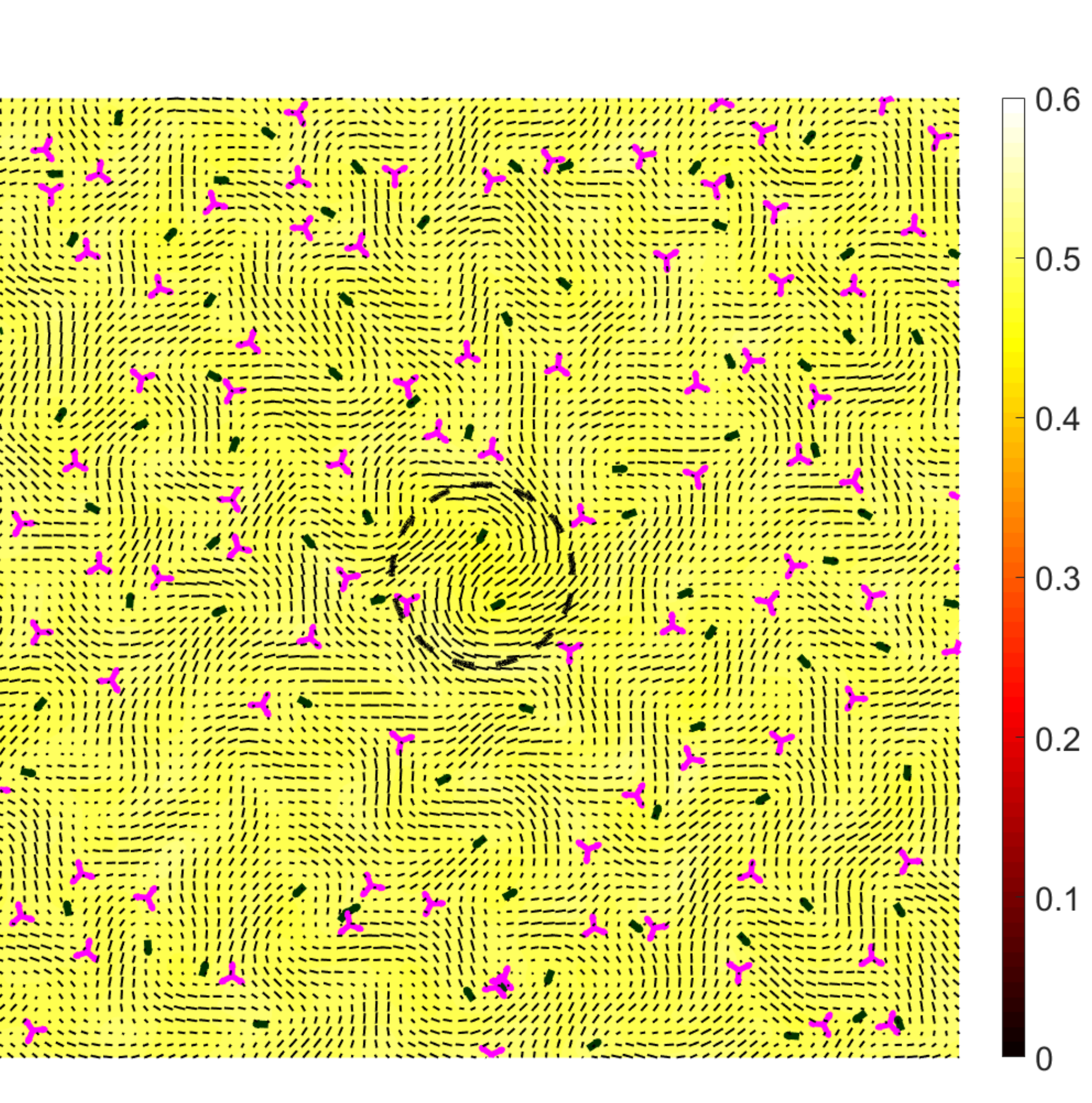}
            \caption{\textbf{Defects can be trapped in the pit area due to the inwards active forces.}
            If the pit radii is sufficiently small compared to the intrinsic active nematic length scale, we find two trapped $+1/2$ circulating around each similar to active nematics confined in circular geometries.} \label{sifig:defect_trapping}
\end{figure}

\subsection*{Movie Captions}

\simov{1}. \textbf{Micropatterned trench geometry.} 
	Fluorescence microscopy video of a bundled microtubule/kinesin network at $1\textrm{mM}$ ATP at the oil-water interface above a submersed SU-8 substrate with a micropatterned trench. 
	The trench possesses a depth $8 \pm 1 \upmu\textrm{m}$ and width $w_t=327 \pm 2 \upmu\textrm{m}$. 
	Scale bar $= 250 \upmu\textrm{m}$. 
	Left and right are shallow regions which exhibit active turbulence, while the centre is the deep region, the edges of which form a well-defined virtual boundary that traps defects.

\simov{2}. \textbf{Simulation results of a repeating lattice of counter-rotating vortices above a trench.}
	The trench width is $w_t = 1.7R$. 
	Plus-half defects (dark green) are trapped between the virtual boundaries, generating the repeating vortex structure along the centre line that is distinct from the active turbulence that exists outside the virtual boundaries. 
	Minus-half defects are shown in magenta. 
	In the colormap, blue (red) denotes clock (anti-clock) wise rotating vortices. 

\simov{3}. \textbf{Micropatterned sinusoid geometry.}
	Fluorescence microscopy video of a bundled microtubule/kinesin active nematic network above an oil-submersed sinusoidal microstructure with a $150\upmu\textrm{m}$ wavelength and a $40\upmu\textrm{m}$ amplitude. 
	The undulations go from left to right such that the parallel troughs and crests run in parallel up and down. 
	Red lines indicate the crest location. Scale bar $= 150 \upmu\textrm{m}$. 
	Train of +1/2 defects move up and down the troughs but are unstable to intermediate pair creation and unbinding events oriented perpendicular to the troughs. 

\simov{4}. \textbf{Micropatterned sinusoid geometry.}
	The same as \mov{mov:undulSmall} but for a microstructure with a $500\upmu\textrm{m}$ wavelength and a $50\upmu\textrm{m}$ amplitude. 
	Red lines indicate the crest location. 
	Scale bar $= 250 \upmu\textrm{m}$.
	Distinct trains of +1/2 are not as immediately apparent, though self-propelled defects tend to align with the troughs. 

\simov{5}. \textbf{Micropatterned stairway geometry.}
	Fluorescence microscopy video of a bundled microtubule/kinesin film over an oil-submersed, micropatterned stairway. 
	Individual steps are $500\upmu\textrm{m}$ wide and $10\upmu\textrm{m}$ tall. 
	The oil depth from left to right deepens from $h-h_0= \{50,60,70,80,90\}\upmu\textrm{m}$ sequentially where the initial fluid depth is $h_0=12\pm3\upmu\textrm{m}$.

\simov{6}. \textbf{Micropatterned pillar geometry.}
	Fluorescence microscopy video of a bundled microtubule/kinesin network at the oil-water interface with $1\textrm{mM}$ ATP above a substrate possessing a micropatterned pillar of radius $r_p=116\pm2\upmu\textrm{m}$ and height $h_p=6.8\pm0.3\upmu\textrm{m}$. 
	The bundled microtubule/kinesin network in the film directly above the immersed pillar is fully depleted. 
	Annihilation events can be observed as +1/2 approach the boundary and annihilate with perimeter-associated -1/2 defects.

\simov{7}. \textbf{Simulation results of the active nematic film above a pillar illustrating depletion of active material.} 
	Active matter concentration $\phi$ (denoted by colormap) is initialised uniformly but eventually depletes from the high friction region inside the circle $(\gamma=0.5)$. 
	The director field is illustrated with black lines and the plus-half defects and minus-half defects are denoted by dark green and magenta circles, respectively.

\simov{8}. \textbf{Micropatterned pillar with +1/2 defect entering depletion zone.} 
	The same as \mov{mov:pillarExp} but exhibiting an instance of active material in the superjacent film crossing the boundary above the pillar. 

\simov{9}. \textbf{Simulation results of defect dynamics around a pillar that acts as a virtual obstacle.} 
	Plus-half defects (dark green) deflect from the depleted high-friction region or are absorbed by minus-half defects (magenta) when approaching the pillar structure. 

\end{document}